\documentclass[12pt]{iopart}
\usepackage{graphicx}
\usepackage{xcolor}

\newcommand{\mc}{\multicolumn{1}{c}}
\begin{document}

\title[Bose-Einstein Condensation on Curved Manifolds]{Bose-Einstein Condensation on Curved Manifolds}

\author{Nat\'alia S. M\'oller}
\address{Physics Department and Research Center Optimas, Technische Universit\"at Kaiserslautern, 67663 Kaiserslautern, Germany, and Departamento de F\'isica, Universidade Federal de Minas Gerais, 31270-901 Belo Horizonte, MG, Brazil}
\ead{natalia.smoller@gmail.com}
\author{F. Ednilson A. dos Santos}
\address{Departamento de F\'isica, Universidade Federal de S\~ao Carlos, 13565-905, S\~ao Carlos, SP, Brazil}
\author{Vanderlei S. Bagnato}
\address{Instituto de F\'isica de S\~ao Carlos, Universidade de S\~ao Paulo, 13560-550 S\~ao Carlos, SP, Brazil, and The Hagler Institute for Advanced Studies at Texas A\&M - USA}
\author{Axel Pelster}
\address{Physics Department and Research Center Optimas, Technische Universit\"at Kaiserslautern, 67663 Kaiserslautern, Germany}
\date{\today}

\vspace{10pt}
\begin{indented}
\item[]August 2020
\end{indented}

\begin{abstract}
Here we describe a weakly interacting Bose gas on a curved smooth manifold, which is embedded in the three-dimensional Euclidean space.~To this end we start by considering a harmonic trap in the normal direction of the manifold, which confines the three-dimensional Bose gas in the vicinity of its surface.~Following the notion of dimensional reduction as outlined in [L.~Salasnich et al., Phys.~Rev.~A {\bf 65}, 043614 (2002)], we assume a large enough trap frequency so that the normal degree of freedom of the condensate wave function can be approximately integrated out. In this way we obtain an effective condensate wave function on the quasi-two-dimensional surface of the curved manifold, where the thickness of the cloud is determined self-consistently. For the particular case when the manifold is a sphere, our equilibrium results show how the chemical potential and the thickness of the cloud increase with the interaction strength.~Furthermore, we determine within a linear stability analysis the low-lying collective excitations together with their eigenfrequencies, which turn out to reveal an instability for attractive interactions.
\end{abstract}

%
%
%
%
%

\section{Introduction}

Bose-Einstein condensates (BECs) are today among the most explored many-body quantum systems. Ranging from more simple thermodynamics~\cite{Stringari,Pethick} to the
more complex turbulent regime~\cite{Turbulence1,Turbulence2,Turbulence3}, BECs represent an interesting workbench for a large variety of both experimental
and theoretical phenomena. In particular, there are many tuning knobs to change the respective system properties in a highly controllable way, among them most importantly its 
dimensionality. Three-dimensional (3D) systems allow to investigate many fundamental macroscopic quantum phenomena as, for instance, superfluidity \cite{Stringari,Pethick,Ueda}.
The two-dimensional (2D) case enables both the formation and the dynamics of vortices or 
the realization of transitions such as the one of the Berezinskii-Kosterlitz-Thouless (BKT) type~\cite{BKT71,BKT73,Zoran}, which is characterized by the binding or unbinding of vortex-antivortex pairs. But so far all experimental confinements of 2D Bose gases took place only in flat space. However it was predicted that interesting topological effects on 2D curved manifolds may occur due to the presence
of vortices and their dynamics~\cite{BKT91,Fetter-Cone,Torus}.

Some time ago it was proposed to use the coupling between the internal atomic structure and external electrodynamic fields in order to  produce a spherical 
shell~\cite{OZobayPRL,OZobayPRA}. With this it is possible to imagine the realization of a BEC on a curved manifold.
Since then there have already been several attempts to confine a Bose gas on the surface of a sphere, which leads to the so-called bubble trap~\cite{HelExp,GarHelReview,GarHelReview2}. However,
due to the gravitational sag, the gas tends to concentrate on the bottom of the sphere. How to realize a Bose gas to occupy the whole sphere is currently an experimental challenge. One way to avoid the effects of
gravity is to perform experiments within microgravity settings~\cite{Micrograv1, Micrograv2, Micrograv3}. The recent installation of the NASA Cold Atom Laboratory (CAL) at the International Space Station (ISS)
nourishes the prospect to realize soon such a bubble trap in microgravity, so that a BEC on a 2D curved manifold will become an experimental reality
\cite{Space1,Space2,Space3}. The planned experimental upgrade BECCAL at the ISS will even allow for a binary Bose mixture of rubidium and potassium to be confined in such a bubble trap~\cite{Space4}.

Inspired by this notion of a bubble trap, several theoretical predictions have already been obtained for both static and dynamic properties of a Bose gas confined on a spherical geometry.
For instance, the collective modes of such a system were already investigated for different regimes \cite{2007}. Furthermore, 
a crossover between 3D and 2D was considered, where the respective limits correspond to a completely filled and a hollow sphere \cite{KSun,KPadavic}.
Additional studies on a sphere deal with the critical temperature for the onset of Bose-Einstein condensation \cite{Vanderlei} and superfluidity
\cite{LSalasnichSphere}. In both cases the limit of an infinitely large radius of the sphere reproduces the corresponding 2D Euclidean results, i.e., in the former case the
Mermin-Wagner-Hohenberg theorem \cite{Mermin,Hohenberg} and in the latter case the BKT phase transition \cite{BKT71,BKT73}.
Moreover, the impact of quantum fluctuations and the formation of clusters were studied~\cite{MonteCarlo}.
Quite recently, even the ground state and collective excitations were analysed for a dipolar Bose-Einstein condensate in a bubble trap \cite{Lima}. Furthermore, the free expansion of a hollow condensate was also studied \cite{LucaMonteCarlo}.

Obviously a sphere does not represent the only possible convenient geometry for confining a Bose gas, therefore a more general consideration of a curved manifold is necessary. For instance, one can use as a starting point
a generic smooth manifold and proceed then with the modelling by taking into account the influence of asymmetries and deformations on the static and thermodynamic properties. 
Following this notion, we restrict ourselves in the present paper to the case of such a smooth manifold. To this end we start in section~\ref{SecPrel} by defining the relevant basic mathematical objects 
such as the Gaussian normal coordinate system to describe a manifold and the Laplace-Beltrami
operator, which extends the Laplacian to a curved manifold. Afterwards, we follow the notion of reference~\cite{LucaSalasnich} to implement a dimensional reduction. Starting from a
3D mean-field description, we introduce in section~\ref{SecNorm} a confining potential, which restricts the condensate in the direction perpendicular to the manifold.
Correspondingly, the 3D wave function factorizes into a 2D part, representing the condensate on the manifold, and a one-dimensional Gaussian function, which describes the confinement. 
With this we derive in section~\ref{SecEnergy} from a 3D Gross-Pitaevskii equation a self-consistent set of equations for both the 2D condensate wave function on the manifold and its width.
Then, section~\ref{SecEquilSphere} discusses the equilibrium case for a sphere.
In section~\ref{SecAction} we formulate the dynamical equations on a generic smooth manifold, while in section~\ref{ColModes} we specialize this dynamics to a sphere and perform subsequently a linear stability analysis. With this we calculate the low-lying frequencies,
which turn out to be stable for repulsive interactions. Finally, section~\ref{Modes} analyses the modes corresponding to these frequencies. It turns out that oscillations with a higher frequency predominantly occur in the
direction of the confinement, whereas oscillations with lower frequencies are mainly restricted on the sphere.

\section{Differential Geometrical Preliminaries} \label{SecPrel}

Due to the fact that in this work we pursue a more general approach to the problem of Bose-Einstein condensation in special geometries, it is necessary to introduce relevant aspects of differential geometry that are used later on. To this end we consider $\mathcal{M}$ to be a smooth manifold~\cite{Lee} to which the Bose gas is confined. The way to describe a manifold is not uniquely defined, i.e., there are a lot of coordinate systems that could be used. Furthermore, it is often not possible to describe the whole manifold with only one coordinate system. In those cases we would have a local coordinate system for each portion of the manifold and we should put all these pieces together to have a global description. The way of choosing these portions and their number depend on the respective manifold and on the chosen coordinate systems. Anyhow, we can assure that this number can be chosen to be finite for compact manifolds, such as spheres and ellipsoids. For not compact manifolds it could be that this number of portions is infinite, but it can always be chosen to be countable. For simplicity, we suppose in the following that the manifold is described by only one coordinate system. Even though it is not true for many examples, it will not affect our main result. Furthermore, all the arguments presented here could be straightforwardly reproduced for the case of needing to describe the manifold with more than one piece.

\begin{figure}[t]
\centering
\includegraphics[scale=0.8]{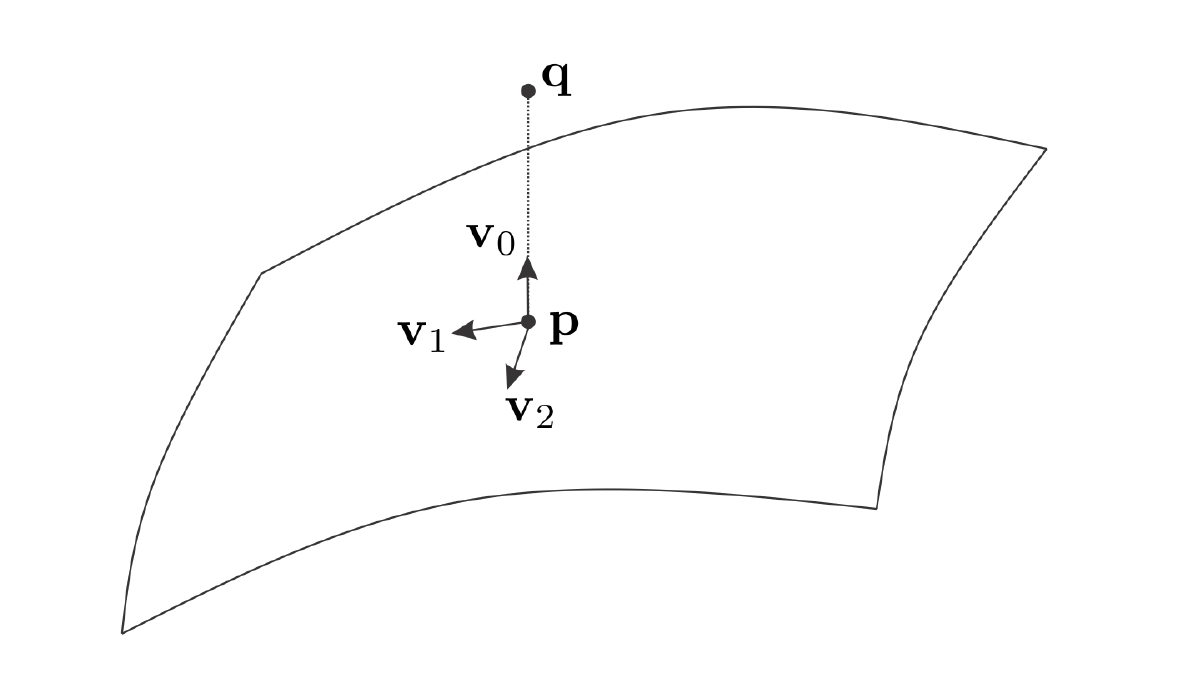}
\caption{Illustration of a portion of the manifold $\mathcal{M}$. The Gaussian normal coordinate system defines the points belonging to this portion via equation (\ref{GaussNorm}) by setting $x^0=0$. The vectors ${\bf v}_1$ and ${\bf v}_2$ are two tangent vectors and ${\bf v}_0(x^1,x^2)$ is a unit normal vector to the manifold at the point ${\bf p}(x^1,x^2)$. The neighbouring point ${\bf q}(x^0,x^1,x^2)$ does not necessarily belong to the manifold $\mathcal{M}$, and the Gaussian normal coordinate system defines this point via equation (\ref{GaussNorm}) by identifying $x^0$ with the distance between ${\bf q}$ and the manifold. In the illustration, ${\bf p}$ represents the point belonging to $\mathcal{M}$, which is the nearest to ${\bf q}$.} \label{FigSurface}
\end{figure}

Here we choose to use a Gaussian normal coordinate system~\cite{GaussCoord,GaussCoordSyst}. It is always possible to do that: any smooth manifold can be locally described by a Gaussian normal coordinate system. According to the illustration in figure~\ref{FigSurface}, these coordinates are defined as follows. Consider a small portion of $\mathcal{M}$, where all the points can be characterized by two real variables $(x^1,x^2)$ belonging to an open subset of the real plane ${\rm I\!R}^2$. So we can denote the points on this manifold by the vector ${\bf p}(x^1,x^2)$. Let ${\bf v}_0(x^1,x^2)$ be a unit normal vector to the manifold at the point ${\bf p}(x^1,x^2)$. Now, we describe the neighbouring points, which do not belong necessarily to $\mathcal{M}$, by
\begin{equation} 
{\bf q}(x^0,x^1,x^2)={\bf p}(x^1,x^2)+x^0{\bf v}_0(x^1,x^2). \label{GaussNorm}
\end{equation}
Note that the points belonging to $\mathcal{M}$ are described by $x^0=0$ and that ones, which do not belong to $\mathcal{M}$, are described by $x^0\neq 0$. Furthermore, fixing any constant value for $x^0$, a locally parallel manifold to $\mathcal{M}$ is defined, which we denote by $\mathcal{M}(x^0)$. In particular, $\mathcal{M}(0)$ simply coincides with the manifold $\mathcal{M}$. 

One detail to be pointed out is that equation~(\ref{GaussNorm}) only represents a local description, so, in principle, $x^0$ cannot be arbitrarily large. We consider that the Gaussian normal coordinate system is well-defined in the interval $|x^0|< R/2$, where $R$ is the minimum, over all points ${\bf p}$ belonging to $\mathcal{M}$, of the smallest curvature radius of each point ${\bf p}$. In order to be more precise, for any fixed point ${\bf p}$ we denote by $R_1({\bf p})$ and $R_2({\bf p})$ the two principal curvature radii of the manifold at that point. Choose as $R({\bf p})$ the minimum between these two radii, i.e.,
\begin{equation}
R({\bf p})=\min\left\{
                 R_1({\bf p}),R_2({\bf p})
               \right\}.
\end{equation}
Then, define $R$ as the minimum value of $R({\bf p})$ over the whole manifold $\mathcal{M}$, i.e.,
\begin{equation}
R=\min_{{\bf p}\in\mathcal{M}}R({\bf p}). \label{R}
\end{equation}
For the special case of a sphere, its radius coincides with $R$. Due to global properties, it could be for some manifold that the Gaussian normal coordinate system describes twice the same point. We exclude such manifolds in the following, restricting ourselves only to manifolds where this situation does not occur.

Now we present a heuristic recipe for deriving equation~(\ref{GaussNorm}). To this end, we suppose to know the manifold equation, which determines the points ${\bf p}$ belonging to a manifold portion as a function of  $x^1$ and $x^2$, that is ${\bf p}={\bf p}(x^1,x^2)$. In order to find the respective tangent vectors we evaluate
\begin{equation}
{\bf v}_1(x^1,x^2)=\frac{\partial {\bf p}(x^1,x^2)}{\partial x^1} \ , \ \ \ \ 
{\bf v}_2(x^1,x^2)=\frac{\partial {\bf p}(x^1,x^2)}{\partial x^2}. \label{TangentV}
\end{equation}
Then the unit normal vector to the manifold at ${\bf p}(x^1,x^2)$ is defined by the cross product between these tangent vectors and a subsequent normalization, i.e.,
\begin{equation}
{\bf v}_0(x^1,x^2)=\frac{{\bf v}_1(x^1,x^2)\times{\bf v}_2(x^1,x^2)}{|{\bf v}_1(x^1,x^2)\times{\bf v}_2(x^1,x^2)|}. \label{DefNormVector}
\end{equation}
With this, equation~(\ref{GaussNorm}) is well-defined and $|x^0|=|({\bf q}-{\bf p})\cdot {\bf v}_0|$ defines the distance of the point ${\bf q}$ from the manifold. 

Now, we can define the tangent vectors to the surface $\mathcal{M}(x^0)$, which is locally parallel to $\mathcal{M}$, using an analogous procedure. From formula~(\ref{GaussNorm}) we calculate the corresponding tangent vectors as ${\bf v}_i(x^0,x^1,x^2)=\partial {\bf q}(x^0,x^1,x^2)/\partial x^i$, for $i=1,2$. The normal vector to the manifold $\mathcal{M}(x^0)$ at the point ${\bf q}(x^0,x^1,x^2)$ is then defined by
\begin{equation}
{\bf v}_0(x^0,x^1,x^2)=\frac{{\bf v}_1(x^0,x^1,x^2)\times{\bf v}_2(x^0,x^1,x^2)}{|{\bf v}_1(x^0,x^1,x^2)\times{\bf v}_2(x^0,x^1,x^2)|} \label{DefNormVectorx0}
\end{equation}
and turns out to be ${\bf v}_0(x^0,x^1,x^2)={\bf v}_0(x^1,x^2)$. The latter statement can be seen from the property ${\bf v}_0(x^1,x^2)\cdot {\bf v}_i(x^0,x^1,x^2)=0$, which follows due to
\begin{equation}
\frac{\partial {\bf v}_0(x^1,x^2)}{\partial x^i}\cdot{\bf v}_0(x^1,x^2)=\frac{1}{2}\frac{\partial[{\bf v}_0(x^1,x^2)\cdot {\bf v}_0(x^1,x^2)]}{\partial x^i}=0,
\end{equation}
for $i=1,2$. With that, a basis for the 3D space at each point ${\bf q}(x^0,x^1,x^2)$ is given by the vectors ${\bf v}_0(x^0,x^1,x^2)$, ${\bf v}_1(x^0,x^1,x^2)$, and ${\bf v}_2(x^0,x^1,x^2)$. This allows us to define a covariant metric of the 3D space in the neighbourhood of the manifold $\mathcal{M}$, since each component of the metric is defined by the scalar product between two respective basis vectors, i.e.,
\begin{equation}
G_{\mu\nu}(x^0,x^1,x^2)={\bf v}_{\mu}(x^0,x^1,x^2)\cdot {\bf v}_{\nu}(x^0,x^1,x^2),
\end{equation}
where $\mu$ and $\nu$ range from $0$ to $2$. From the definition of the normal vector ${\bf v}_0$ in equation~(\ref{DefNormVector}) and from the above discussion, we conclude both $G_{00}(x^0,x^1,x^2)={\bf v}_{0}(x^1,x^2)\cdot {\bf v}_{0}(x^1,x^2)=1$ and $G_{0i}(x^0,x^1,x^2)=G_{i0}(x^0,x^1,x^2)={\bf v}_{0}(x^1,x^2)\cdot {\bf v}_{i}(x^0,x^1,x^2)=0$, for $i=1,2$, for all $x^0$, $x^1$ and $x^2$ where the coordinate system is well-defined. With that and using the Gaussian normal coordinate system, we obtain that the covariant metric for the 3D space in the surrounding of the manifold $\mathcal{M}$ can be represented by the matrix
\begin{equation}G_{\mu\nu}(x^0,x^1,x^2)= 
\renewcommand{\arraystretch}{1.2}
  \left(
  \begin{array}{ c | c c }
    \mc{1} & \ \ \ \ \ \ 0 & \ 0 \\
    \cline{2-3}
    0  & &  \\
    0  & \multicolumn{2}{c}{\raisebox{.6\normalbaselineskip}[0pt][0pt]{$g_{ij}(x^0,x^1,x^2)$}} \\
  \end{array}
  \right), \label{metric}
\end{equation}
for $\mu$ and $\nu$ ranging from $0$ to $2$, while $i$ and $j$ range only from $1$ to $2$.
For $x^0$ fixed, $g_{ij}(x^0,x^1,x^2)$ denotes the covariant metric of the manifold $\mathcal{M}(x^0)$. Each entry of this metric represents the scalar product of the respective tangent vectors ${\bf v}_1(x^0,x^1,x^2)$ and ${\bf v}_2(x^0,x^1,x^2)$. For the special case $x^0=0$, $g_{ij}(0,x^1,x^2)$ represents the metric of the manifold $\mathcal{M}$ and we use the abbreviated notation $g_{ij}(x^1,x^2)$.

Note that the metric $g_{ij}(x^0,x^1,x^2)$ can be Taylor expanded around the metric $g_{ij}(x^1,x^2)$ and written in terms of other local properties of the manifold $\mathcal{M}$, as is summarized in 
appendix A. There, it is also shown that calculating the square root of the determinant of this metric yields
\begin{equation}
\sqrt{\det g_{ij}(x^0,x^1,x^2)}=\sqrt{(\det g)}\cdot
 \left[
   1+\mathcal{O}
     \left(
       \frac{x^0}{R}
     \right)
   +\mathcal{O}
     \left(
       \frac{(x^0)^2}{R^2}
     \right)+...
 \right], \label{Taylor}
\end{equation}
for $|x^0|<R/2$, where we have introduced the notation $\det g=\det g_{ij}(x^1,x^2)$. This result turns out to be useful in the next sections. Note that we can assume in the following without loss of generality that singular points, where $\det g$ vanishes, do not occur. One possibility would be the presence of a coordinate singularity, which occurs, for instance, at the north and south poles of a sphere, when a spherical coordinate system is used. But, as we comment explicitly below, such coordinate singularities at the poles of a sphere turn out to have no physical consequence. Another possibility would be a non-coordinate system singularity, which includes, for instance, regions similar to the edge of a cone. But such manifolds with a non-coordinate system singularity are discarded from our approach as they would correspond to a not smooth manifold. Basically, such a real singularity in the metric involves a drastic change of the geometry, which has to be analysed case by case in more detail within another study.

We know that in many-body quantum theory the kinetic energy of a 3D BEC is given by the Laplacian of the condensate wave function. A generalization of the Laplacian for the 3D space in a generalized coordinate system is called Laplace-Beltrami operator~\cite{Jurgen}. It is expressed by
\begin{equation}
\Delta_{\rm LB}=\frac{1}{\sqrt{\det G \ }}\frac{\partial}{\partial x^\eta}\left(\sqrt{\det G \ } G^{\eta\kappa}\frac{\partial}{\partial x^\kappa}
\right), \label{LB}
\end{equation}
with $\eta$ and $\kappa$ ranging from $0$ to $2$ and $\det G$ denoting the determinant of the covariant metric. Note that we use the Einstein notation in (\ref{LB}), so that a summation over the same co- and contravariant indices is implicitly assumed. For a flat 3D space described by a Cartesian coordinate system, the metric is given by an Euclidean metric with the Kronecker symbol as its respective components and the above formula recovers the standard representation of the Laplacian. On the other side, still for the flat 3D space represented via the Gaussian normal coordinate system, using the metric~(\ref{metric}), the Laplace-Beltrami operator~(\ref{LB}) reduces to
\begin{equation}
\Delta=\frac{\partial^2}{\partial {x^0}^2}+
\frac{\partial}{\partial x^0}\left(\ln\sqrt{\det g \ }\right) \frac{\partial}{\partial x^0}+\Delta_{\mathcal{M}(x^0)},\label{LaplacCurv1}
\end{equation}
with the abbreviation
\begin{equation}
\Delta_{\mathcal{M}(x^0)}=\frac{1}{\sqrt{\det g \ }}\frac{\partial}{\partial x^i}\left(\sqrt{\det g \ } g^{ij}\frac{\partial}{\partial x^j}
\right). \label{DeltaM}
\end{equation}
Note that~(\ref{DeltaM}) denotes also a Laplace-Beltrami operator, but this time in the context of the manifold $\mathcal{M}(x^0)$ for each fixed value of $x^0$, and the indices $i,j$ range from $1$ to $2$. 
In particular, when $x^0=0$, the operator~(\ref{DeltaM}) represents the Laplace-Beltrami operator of the manifold $\mathcal{M}$, denoted in the following simply by $\Delta_{\mathcal{M}}$.

\section{Normalization of Condensate Wave Function} \label{SecNorm}

Now that we have our mathematical objects well-defined, we introduce two important physical quantities, which we need to consider when dealing with a BEC on a manifold: the potential which confines the Bose gas to the manifold and a particular ansatz for the condensate wave function.

We suppose that the Bose gas is confined in the immediate vicinity of the manifold. Such a confinement could be realized, for instance, by a harmonic oscillator potential in the normal direction to the manifold, which has its minimum on the manifold. In the Gaussian normal coordinate system introduced in the previous section, this potential has the form
\begin{equation}
V_{\rm harm}(x^0)=\frac{1}{2}M\omega^2(x^0)^2. \label{potential}
\end{equation}
Here the particle mass $M$ and the frequency $\omega$ define a length scale in terms of the oscillator length $\sigma_{\rm osc}=\sqrt{\hbar/M\omega}$, which represents the order of magnitude of the thickness of the Bose gas cloud surrounding the manifold. In the following we assume that the frequency $\omega$ is so large that the oscillator length $\sigma_{\rm osc}$ is much smaller than the minimum $R$ of the respective local curvature radii on the manifold, i.e., $\sigma_{\rm osc}\ll R$. This corresponds to the physical situation that the Bose gas forms a thin shell around the manifold.

In addition to this harmonic potential, we also allow the Bose gas to be affected by a potential $U$ depending on the manifold coordinates $x^1$ and $x^2$. Thus, the total potential is given by
\begin{equation}
V(x^0,x^1,x^2)=V_{\rm harm}(x^0)+U(x^1,x^2). \label{3dpotential}
\end{equation}
Note that $V_{\rm harm}(x^0)$ and $U(x^1,x^2)$ are 3D potentials, even though each one does not depend on all three variables.

Let us now compute as a physical quantity the number of particles. Denoting the condensate wave function $\Psi(x^0,x^1,x^2)$, the number of particles is given by an integral over the whole 3D space of its squared norm, i.e., $N=\int dV |\Psi(x^0,x^1,x^2)|^2$. But since the gas is confined in the vicinity of the manifold $\mathcal{M}$, the integral becomes naturally restricted to the manifold neighbourhood $\mathcal{N(M)}$, yelding $N=\int_{\mathcal{N(M)}} dV |\Psi(x^0,x^1,x^2)|^2$. We consider that this neighbourhood is defined by the points ${\bf q}(x^0,x^1,x^2)$ of the 3D space described by the Gaussian coordinate system according to equation~(\ref{GaussNorm}) with $|x^0|< R/2$, with $R\gg\sigma_{\rm osc}$.
Thus, the number of particles in the gas can be expressed as
\begin{eqnarray}
N=\int_{-R/2}^{R/2}dx^0\int dx^1dx^2\sqrt{\det g(x^0,x^1,x^2)}|\Psi(x^0,x^1,x^2)|^2. \label{NInt}
\end{eqnarray}

In order to describe the confinement of the Bose gas in a thin shell, one is tempted to follow the arguments of the reference~\cite{LucaSalasnich} and choose a trial wave function of the form
\begin{equation}
\Psi(x^0,x^1,x^2)=\frac{e^{\frac{-(x^0)^2}{2\sigma(x^1,x^2)^2}}}{\sqrt[4]{\pi}\sqrt{\sigma(x^1,x^2)}}\cdot\psi (x^1,x^2). \label{wrongtrial}
\end{equation}
We plug this trial function into the above normalization integral~(\ref{NInt}) and expand the term $\sqrt{\det g(x^0,x^1,x^2)}$ in a power series using equation~(\ref{Taylor}). After this expansion, we are able to approximately perform the integrals in the limit of large $R$, which leads to an exponentially small error $\mathcal{O}(e^{-R^2/\sigma^2})$~\cite[(8.25)]{Gradshteyn}. The integral over $x^0$ of each term of the power series times a Gaussian is given by
\begin{equation}
\int_{-\infty}^\infty dx^0\frac{(x^0)^{n}e^{-(x^0)^2/\sigma^2}}{\sqrt{\pi}\sigma}=
\frac{(n-1)!!\sigma^{n}}{2^{n/2}},
\end{equation}
for even non-negative integer values of $n$, while it vanishes for odd non-negative values of $n$. Therefore, only the even order terms survive and provide results at least of the order of $\sigma^2/R^2$. Thus, the number of particles can be calculated through the normalization of the 2D function $\psi(x^1,x^2)$ apart from a polynomial error, i.e.,
\begin{equation}
N=\int dx^1dx^2\sqrt{\det g(0,x^1,x^2)} |\psi(x^1,x^2)|^2+\mathcal{O}\left(\frac{\sigma^2}{R^2}\right). \label{NormWrong}
\end{equation} 
But now we argue that a better choice of the trial wave function is provided by
\begin{equation}
\Psi(x^0,x^1,x^2)=\frac{e^{\frac{-(x^0)^2}{2\sigma(x^1,x^2)^2}}}{\sqrt[4]{\pi}\sqrt{\sigma(x^1,x^2)}}\cdot \psi(x^0,x^1,x^2), \label{trial}
\end{equation}
where we have defined
\begin{equation}
\psi(x^0,x^1,x^2)=\frac{\phi(x^1,x^2)}{\sqrt[4]{\det g(x^0,x^1,x^2)}}.\label{trialg}
\end{equation}
For $x^0=0$ we have that $\psi(0,x^1,x^2)$ represents the 2D wave function of the gas. In order to calculate the number of particles, we can follow the same procedure as the one above, but now we do not need to perform a Taylor series expansion, since the denominator of the term~(\ref{trialg}) matches with the one from the volume element. To perform the integral it is only necessary to approximately take the limit of large $R$, leading to an exponentially small error.
Thus, the number of particles is given by the integral of the squared norm of the 2D wave function apart from only an exponentially small error:
\begin{equation}
N=\int dx^1dx^2\sqrt{\det g(0,x^1,x^2)} |\psi(0,x^1,x^2)|^2+\mathcal{O}(e^{-R^2/\sigma^2}). \label{NormGood}
\end{equation}
We conclude that the ansatz~(\ref{trial}), (\ref{trialg}) is a much better approximation than~(\ref{wrongtrial}), as the normalization of the wave function in~(\ref{NormGood}) is more accurate than~(\ref{NormWrong}). Therefore, we investigate in the following the consequences of~(\ref{trial}), (\ref{trialg}) in view of reducing the original 3D problem to an effective 2D one.

Note that in section~\ref{SecPrel} we have commented about singularities at the poles of a sphere when we use spherical coordinates and this could make equation~(\ref{trialg}) not well-defined at these points. However, this turns out not to be the case, since the term $\phi(x^1,x^2)$
would also vanish at the poles, thus compensating the coordinate singularities.

\section{Reducing Dimensionality} \label{SecEnergy}

In the previous sections we introduced the necessary differential geometrical notation and the physical ideas on how to confine a weakly interacting Bose-Einstein condensate in the neighbourhood of a manifold. Now we proceed with the description of this system by considering its grand-canonical energy
\begin{equation}
E=\int dV\Psi^*
\left(
-\frac{\hbar^2}{2M}\Delta+\frac{M\omega^2}{2}(x^0)^2+U(x^1,x^2)+\frac{1}{2}g_{\rm int}|\Psi|^2-\mu
\right)
\Psi, \label{Energy}
\end{equation}
where $g_{{\rm int}}=4\pi \hbar^2 a_s/M$ denotes the interaction strength, determined by the $s$-wave scattering length $a_s$, and $\mu$ is the chemical potential of the system.
Since the confinement frequency $\omega$ is supposed to be large enough, we can follow the same procedure as in the last section and perform this integral only on the manifold neighbourhood $\mathcal{N(M)}$. Then, the energy~(\ref{Energy}) is well approximated by
\begin{eqnarray}
E=\int_{\mathcal{N(M)}} dx^1dx^2dx^0\sqrt{\det g \ } \ \Psi^*  \label{EnergyFunctional}
\\ \hspace{1.8cm}\cdot \Bigg(
-\frac{\hbar^2}{2M}\Delta  +\frac{M\omega^2}{2}(x^0)^2+U(x^1,x^2)+\frac{1}{2}g_{\rm int}|\Psi|^2-\mu
\Bigg)
\Psi. \nonumber 
\end{eqnarray}
Inserting the trial function~(\ref{trial}), (\ref{trialg}) into the energy functional~(\ref{EnergyFunctional}), we expand it into a Taylor series with respect to $x^0$ and perform the resulting integral with respect to $x^0$ approximately in the limit of large $R$ as in section~\ref{SecNorm}.

Note that the Laplacian~(\ref{LaplacCurv1}) with the trial function (\ref{trial}), (\ref{trialg}) reads explicitly
\begin{eqnarray}
\hspace{-1cm} \Delta\Psi=
\Bigg\{
   \Bigg[
       \left(
          \frac{(x^0)^2}{\sigma^4}-\frac{1}{\sigma^2}
       \right)
     -\frac{1}{4}
        \left(
           \frac{\partial\ln\sqrt{\det g \ }}{\partial x^0}
        \right)^2
    -\frac{1}{2}\frac{\partial^2\ln\sqrt{\det g \ }}{\partial {x^0}^2}
 \\
\hspace{1cm} 
 +\frac{g^{ij}(\partial_i{\sigma})(\partial_j{\sigma})}{2\sigma^ 2}\Bigg]
 \psi(x^0,x^1,x^2)
 +\Delta_{\mathcal{M}(x^0)}\psi(x^0,x^1,x^2)
\Bigg\} 
\frac{e^{\frac{-(x^0)^2}{2\sigma^2}}}{\sqrt[4]{\pi}\sqrt{\sigma}}.\nonumber
\end{eqnarray}
Approximately integrating the energy~(\ref{EnergyFunctional}) with respect to $x^0$, we obtain
\begin{eqnarray}
\hspace{-2.5cm} E=\int_{\mathcal{N(M)}} dx^1dx^2 \sqrt{\det g \ } \ \psi^*
    \Bigg\{
     -\frac{\hbar^2}{2M}\Delta_{\mathcal{M}(x^0)}
     -\frac{\hbar^2g^{ij}}{4M}\frac{(\partial_i{\sigma})(\partial_j{\sigma})}{\sigma^ 2}
-\mu
  \label{Energyx1x2}
\\ 
\hspace{-2.5cm}+\frac{\hbar^2}{4M}\Bigg[
      \frac{1}{2}
          \left(
             \frac{\partial\ln\sqrt{\det g \ }}{\partial x^0}
          \right)^2
        +\frac{\partial^2\ln\sqrt{\det g \ }}{\partial {x^0}^2}
    \Bigg]
+\frac{\hbar^2}{4M\sigma^2}
+\frac{M\omega^2\sigma^2}{4}
+U
+\frac{g_{\rm int}|\psi|^2}{2\sqrt{2\pi} \ \sigma}
    \Bigg\}\psi\Bigg|_{x^0=0}. \nonumber
\end{eqnarray}
Based on the ideas of reference~\cite{LucaSalasnich}, we extremize the resulting energy~(\ref{Energyx1x2}) with respect to both $\psi^*$ and $\sigma$.  In the first case we obtain the time-independent 2D Gross-Pitaevskii equation 
\begin{eqnarray}
\hspace{-2.5cm}     \mu\psi=\left(
                -\frac{\hbar^2}{2M}\Delta_{\mathcal{M}}
                -\frac{\hbar^2g^{ij}}{4M}\frac{(\partial_i{\sigma})(\partial_j{\sigma})}{\sigma^ 2}
                +V_{\rm eff}
                +U
                +\frac{\hbar^2}{4M\sigma^2}
                +\frac{M\omega^2\sigma^2}{4}
                +g_{\rm 2D}|\psi|^2
            \right) \psi, \label{2DGPpsi}
\end{eqnarray}
where $\psi=\psi(x^0=0,x^1,x^2)$ denotes the 2D wave function, while $\Delta_\mathcal{M}$ is given in equation~(\ref{DeltaM}) in the case when $x^0=0$, and
\begin{equation}
V_{\rm eff}(x^1,x^2)=\frac{\hbar^2}{4M}\Bigg[
      \frac{1}{2}
          \left(
             \frac{\partial\ln\sqrt{\det g \ }}{\partial x^0}
          \right)^2
        +\frac{\partial^2\ln\sqrt{\det g \ }}{\partial {x^0}^2}
    \Bigg]\Bigg|_{x^0=0}+... \label{Veff}
\end{equation}
represents an effective potential due to the non-trivial metric. Moreover, $g_{\rm 2D}=g_{\rm int}/(\sigma\sqrt{2\pi})$ turns out to be the 2D interaction parameter, which gets larger for smaller values of the width $\sigma$. It shows that a strong confinement leads to stronger effective two-particle interactions on the manifold.

Extremizing instead the energy~(\ref{Energyx1x2}) with respect to the cloud width $\sigma$ yields
\begin{eqnarray}
\frac{\hbar^2}{2M\sigma^3}
-\frac{M\omega^2\sigma}{2}
+\frac{g_{\rm int}|\psi|^2}{2\sqrt{2\pi} \sigma^2}
+\frac{\hbar^2}{2M}\frac{(\Delta_{\mathcal{M}}\sigma)}{\sigma^2} \nonumber \\
-\frac{\hbar^2}{2M}g^ {ij}\frac{(\partial_i \sigma)(\partial_j \sigma)}{\sigma^3} 
+\frac{\hbar^2}{2M}\frac{g^{ij}(\partial_i \sigma)}{\sigma^2}\frac{(\partial_j |\psi|^ 2)}{|\psi|^ 2}=0. \label{2DGPpsisigma}
\end{eqnarray}
Thus, in equilibrium, one has to solve both equations~(\ref{2DGPpsi}) and (\ref{2DGPpsisigma}) for $\psi$ and $\sigma$ by taking into account the particle number in equation~(\ref{NormGood}).
Note that our results~(\ref{2DGPpsi}) and (\ref{2DGPpsisigma}) for a general manifold contain the corresponding ones for a plane, which were already treated in reference~\cite{LucaSalasnich} for a constant width $\sigma$.

\section{Equilibrium on a Sphere} \label{SecEquilSphere}

The simplest case to be studied is the ground state of a sphere with radius $R$. For simplicity, we suppose that we do not have any external potential, i.e., $U(x^1,x^2)=0$. Furthermore, from~(\ref{Veff}) we conclude that the effective potential $V_{\rm eff}$ vanishes for such a sphere. Due to the rotational symmetry of the sphere, the gas in the ground state is described by a uniform distribution, thus the 2D wave function $\psi_0$ satisfies $\Delta_\mathcal{M}\psi_0=0$. Moreover, from the normalization~(\ref{NormGood}) we obtain
\begin{equation}
\psi_0^2=\frac{N}{4\pi R^2}. \label{psinot}
\end{equation}
With this the time-independent 2D Gross-Pitaevskii equation~(\ref{2DGPpsi}) reduces to an algebraic equation for the chemical potential~$\mu$, i.e., the equation of state
\begin{eqnarray}
\frac{\mu}{\hbar\omega}=
\frac{1}{4}
    \left(
       \frac{\sigma_{\rm osc}^2}{\sigma_0^2}
      +\frac{\sigma_0^2}{\sigma_{\rm osc}^2}
    \right)
+P\frac{\sigma_{\rm osc}}{\sigma_0}, \label{RelationMiP}
\end{eqnarray}
where
\begin{equation}
P=\frac{a_s\sigma_{\rm osc}N}{\sqrt{2\pi} \ R^2}, \label{DimensionlessInt}
\end{equation}
represents the dimensionless interaction strength. Note that $P$ can be tuned by changing the particle number $N$, the $s$-wave scattering length $a_s$, as well as by changing the oscillator length $\sigma_{\rm osc}$ or the radius $R$.

\begin{center}
\begin{figure}[t]
\includegraphics[scale=0.5]{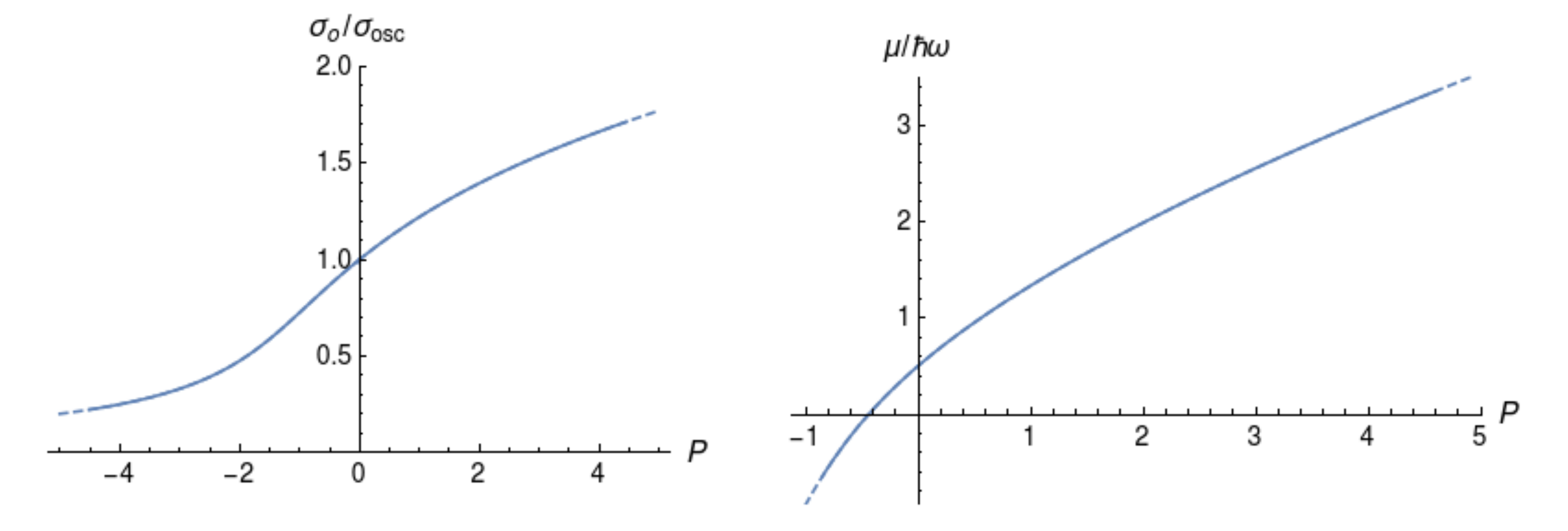}
\caption{Dimensionless width $\sigma_0/\sigma_{\rm osc}$ and chemical potential $\mu/\hbar\omega$ as functions of dimensionless interaction strength $P$.}\label{EquilibriumGraphics}
\end{figure}
\end{center}

Correspondingly, for a sphere also the ground state thickness $\sigma_0$ is uniform, so~(\ref{2DGPpsisigma}) reduces to
\begin{equation}
\frac{\sigma_0^4}{\sigma^4_{\rm osc}}=1+P\frac{\sigma_0}{\sigma_{\rm osc}}. \label{PolynSigmaP}
\end{equation}
In figure~\ref{EquilibriumGraphics} we plot the results for the dimensionless Gaussian width $\sigma_0/\sigma_{\rm osc}$ and the dimensionless chemical potential $\mu/\hbar\omega$ as functions of the dimensionless interaction strength $P$.

The width of the Gaussian is positive for any value of $P$. It coincides with the harmonic oscillator length $\sigma_{\rm osc}$ for vanishing interactions and increases for repulsive interaction strengths ($P>0$). For strong repulsive interactions, its asymptotic behaviour is of the form $\sigma_0/\sigma_{\rm osc}=\sqrt[3]{P}$. For attractive interaction strengths ($P<0$) it decreases as $|P|$ increases, tending to zero with an asymptotic behaviour of the form $\sigma_0/\sigma_{\rm osc}=-1/P$. 

The dimensionless chemical potential coincides with $1/2$ for vanishing interactions and increases for positive values of $P$. Its asymptotic behaviour for large positive values of $P$ is $\mu/\hbar\omega=(5/4)P^{2/3}$. It reaches zero at about $P\approx -0.44$ and turns to be negative for smaller values of $P$. Its asymptotic behaviour for large negative values of $P$ is given by $\mu/\hbar\omega=-3P^{2}/4$.

In order to have some intuitive notion of these dimensionless values, suppose that one is arranging to perform a bubble trap experiment in microgravity with about $N=10^5$ rubidium atoms on a sphere with a radius about $R=10 \ \mu$m and a harmonic trap such that the harmonic oscillator length is of the order of 
$\sigma_{\rm osc}=1 \ \mu$m~\cite{PrivHelPer}. As the $s$-wave scattering length $a_s$ is about $100$ times the Bohr radius, we obtain a dimensionless interaction of about $P=2.1$. For later figures we always use those experimentally realistic parameters.

Note that our equilibrium results recover the corresponding ones for an infinite plane in the limit of an infinite curvature radius, i.e., $R\rightarrow\infty$. To this end we just have to redefine equations~(\ref{psinot}) and (\ref{DimensionlessInt}) as $\psi_0^2=\rho$ and $P=2\sqrt{2\pi} a_s\sigma_{\rm osc}\rho$ by introducing the particle density $\rho$ of the plane.

\section{Dynamics on a Curved Manifold} \label{SecAction}

In the last section, we calculated equilibrium results for a gas confined on a sphere, finding that the cloud has a positive width for all interactions strengths. In order to determine the stability of the system, we now embark upon a linear stability analysis. To this end, we extend the equilibrium consideration of section~\ref{SecEnergy} and treat the Bose gas dynamically.

We begin a dynamical analysis deriving the temporal evolution of the 2D wave function and of the cloud width. To do that, instead of the energy, we use the corresponding action
\begin{equation}
\hspace{-1cm} S=\int dt\int dV\Psi^*
\left(
i\hbar\partial_t
+\frac{\hbar^2}{2M}\Delta
-\frac{M\omega^2}{2}(x^0)^2
-U(x^1,x^2)
-\frac{1}{2}g_{\rm int}|\Psi|^2
\right)
\Psi, \label{Action}
\end{equation}
and a more general ansatz than that of equation~(\ref{trial}), which includes an imaginary width term in the Gaussian exponent, according to \cite{PRL1996,PRA1997}
\begin{equation}
\Psi(x^0,x^1,x^2,t)=\frac{\exp\left[\frac{-(x^0)^2}{2}\left(\frac{1}{\sigma(x^1,x^2,t)^2}+iB(x^1,x^2,t)\right)\right]}{\sqrt[4]{\pi}\sqrt{\sigma(x^1,x^2,t)}}\cdot  \psi(x^0,x^1,x^2,t) \ , \label{AnsatzB}
\end{equation}
with
\begin{equation}
\psi(x^0,x^1,x^2,t)=\frac{\phi(x^1,x^2,t)}{\sqrt[4]{\det g(x^0,x^1,x^2)}}. \label{raizgt}
\end{equation}
Here $B$ represents the variational parameter conjugated to the cloud width, which is necessary to be included in order to properly describe the dynamics of the system.

Using the same procedure as in the last section, we insert this ansatz into equation~(\ref{Action}) and integrate approximately the variable $x^0$. With this we get the following expression for the action:
\begin{eqnarray}
\nonumber
\hspace{-2.5cm} S=\int_{\mathcal{N(M)}} dtdx^1dx^2 \sqrt{g}\psi^*
     \Bigg[
     i\hbar\partial_t\psi
     +\frac{\hbar^2}{2M}(\Delta_{\mathcal{M}(x^0)}\psi)  
  +\frac{\hbar\sigma^2}{4}(\partial_t B)\psi
  -\frac{\hbar^2}{4M}\sigma^2B^2\psi
\\ 
\nonumber
\hspace{-2.2cm}
-\frac{\hbar^2}{2M}g^{ij}
   \left(
     \frac{(\partial_i\sigma)(\partial_j\sigma)}{2\sigma^2}\psi
     +\frac{3\sigma^4}{16}(\partial_iB)(\partial_jB)\psi
     +\frac{i\sigma}{2}(\partial_i\sigma)(\partial_jB)\psi
     +\frac{i\sigma^2}{2}(\partial_iB)(\partial_j\psi)
   \right)
   \\
\hspace{-2.2cm}  -\frac{i\hbar^2\sigma^2}{8M}(\Delta_\mathcal{M}B)\psi
-V_{\rm eff}\psi
-\frac{\hbar^2}{4M\sigma^2}\psi
-\frac{M\omega^2\sigma^2}{4}\psi
-\frac{g_{\rm int}|\psi|^2}{2\sqrt{2\pi}\sigma}\psi
\Bigg]\Bigg|_{x^0=0}. \label{eq37}
\end{eqnarray}
Following the standard approach, we extremize the action~(\ref{eq37}) with respect to $\psi^*$, $\sigma$ and $B$. In this way, we obtain at first the evolution equation for $\psi$
\begin{eqnarray}
\nonumber    
\hspace{-2.5cm}     i\hbar\partial_t\psi =
     -\frac{\hbar^2}{2M}(\Delta_{\mathcal{M}}\psi)  
  -\frac{\hbar\sigma^2}{4}(\partial_t B)\psi
  +\frac{\hbar^2}{4M}\sigma^2B^2\psi
  +\frac{i\hbar^2\sigma^2}{8M}(\Delta_{\mathcal{M}}B)\psi
\\
\nonumber
\hspace{-1.5cm}
+\frac{\hbar^2}{2M}g^{ij}
   \left(
     \frac{(\partial_i\sigma)(\partial_j\sigma)}{2\sigma^2}\psi
     +\frac{3\sigma^4}{16}(\partial_iB)(\partial_jB)\psi
     +\frac{i\sigma}{2}(\partial_i\sigma)(\partial_jB)\psi
     +\frac{i\sigma^2}{2}(\partial_iB)(\partial_j\psi)
   \right)
\\ 
\hspace{-1.5cm}
+V_{\rm eff}\psi
+\frac{\hbar^2}{4M\sigma^2}\psi
+\frac{M\omega^2\sigma^2}{4}\psi
+\frac{g_{\rm int}|\psi|^2}{\sqrt{2\pi}\sigma}\psi, \label{PsiComplicado}
\end{eqnarray}
as well as the corresponding equation for $\sigma$
\begin{eqnarray}
\nonumber
\frac{M^2\omega^2}{\hbar^2}\sigma^4
&=1+\frac{g_{\rm int}M\sigma|\psi|^2}{\hbar^2\sqrt{2\pi}}
+\sigma\Delta_{\mathcal{M}}\sigma
+\frac{M\sigma^4}{\hbar}\partial_t B
\\ \label{SigmaComplicado}
&-\sigma^4B^2
-g^{ij}(\partial_i\sigma)(\partial_j\sigma)
+g^{ij}\sigma(\partial_i\sigma)\frac{\partial_j|\psi|^2}{|\psi|^2}
\\
\nonumber
&-\frac{3}{4}\sigma^6g^{ij}(\partial_iB)(\partial_jB)
+\frac{i}{2}\sigma^4g^{ij}(\partial_iB)(\psi\partial_j\psi^*-\psi^*\partial_j\psi).
\end{eqnarray}
A subsequent extremization of the action with respect to $B$ yields
\begin{eqnarray}
\hspace{-1.5cm} B=-\frac{M}{\hbar}\frac{\partial_t\sigma}{\sigma}
-\frac{M}{2\hbar}\frac{\partial_t|\psi|^2}{|\psi|^2}
+\frac{3\sigma^2}{8}\Delta_{\mathcal{M}}B
+\frac{i}{4}\left(
               \frac{\Delta_{\mathcal{M}}\psi}{\psi}
              -\frac{\Delta_{\mathcal{M}}\psi^*}{\psi^*}
            \right) \label{BComplicado}
\\
\nonumber
\hspace{-1cm}+\frac{i}{2}g^{ij}\frac{\partial_i\sigma}{\sigma}
  \left(
     \frac{\partial_j\psi}{\psi}
     -\frac{\partial_j\psi^*}{\psi^*}
  \right)
+\frac{3\sigma^2}{8}g^{ij}(\partial_iB)\frac{\partial_j|\psi|^2}{|\psi|^2}
+\frac{3\sigma^2}{2}g^{ij}\frac{\partial_i\sigma}{\sigma}\partial_jB.
\end{eqnarray}
We remark that in the above three equations~(\ref{PsiComplicado})--(\ref{BComplicado}), the value of the first variable is fixed $x^0=0$, such that $\psi$ stands for $\psi(0,x^1,x^2,t)$. These equations describe the dynamics of a Bose gas on a curved manifold by determining the evolution of the 2D wave function, as well as the real and imaginary cloud width self-consistently.

\section{Collective Modes of BEC on a Sphere} \label{ColModes}

Collective modes of a condensate are of great value for experimental studies, since they allow a quantitative characterization of the underlying system, even when an optical absorption projection is made in the data collection. In addition, collective modes are associated with the equilibrium state around which they occur. Being able to analyse the collective modes of a confined condensate creates the possibility of understanding the influence of various system parameters on the hydrodynamics of the system.

In order to determine the low-lying collective modes, we now study small perturbations of the ground state for a Bose gas confined on the surface of a sphere of radius $R$. To this end, we perform a linear stability analysis of the evolution equations derived in the previous section. Note that the effective potential $V_{\rm eff}(x^1,x^2)$ in equation~(\ref{PsiComplicado}) vanishes for the case of a sphere. We suppose a small perturbation of the ground state in the form
\begin{eqnarray}
\psi&=(\psi_0+\delta\psi)e^{-i\mu t/ \hbar} \label{smallpsidimension}
\\
\sigma&=\sigma_0+\delta\sigma \label{smallsigma}
\\
B&=\delta B, \label{smallB}
\end{eqnarray}
where $\psi_0$ is given in equation~(\ref{psinot}), the chemical potential $\mu$ follows from equation~(\ref{RelationMiP}), and $\sigma_0$ is defined via equation~(\ref{PolynSigmaP}).

Inserting the perturbed quantities~(\ref{smallpsidimension})--(\ref{smallB}) into equations~(\ref{PsiComplicado})--(\ref{BComplicado}) and considering only the first order terms of $\delta\psi$, $\delta\sigma$ and $\delta B$, we obtain
\begin{eqnarray}
\hspace{-2.5cm}
i\hbar(\partial_t\delta\psi)
+\frac{\hbar\sigma_0^2}{4}(\partial_t \delta B)=
\frac{1}{2MR^2}(L^2\delta\psi)
+g_{\rm 2D}\psi_0^2(\delta\psi+\delta\psi^*)
-\frac{i\sigma_0^2}{8\hbar MR^2}(L^2 \delta B) \label{difeq1}
\\
\hspace{4.3cm}
+\left(
  -\frac{\hbar^2}{2M\sigma_0^3}
  +\frac{M\omega^2\sigma_0}{2}
  -\frac{g_{\rm 2D}\psi_0^2}{\sigma_0}
\right)
\psi_0\delta\sigma, \nonumber
\end{eqnarray}
\begin{eqnarray}
\hspace{-2.5cm}\frac{M\sigma_0^4}{\hbar}\partial_t \delta B=
-\frac{g_{\rm 2D}M\sigma_0^2\psi_0}{\hbar^2}(\delta\psi+\delta\psi^*)
+\left(
  \frac{\sigma_0}{\hbar^2R^2}L^2
  +\frac{4M\omega^2\sigma_0^3}{\hbar^2}
  -\frac{g_{\rm 2D}M\sigma_0\psi_0^2}{\hbar^2}
\right)
\delta\sigma, \label{difeq2}
\end{eqnarray}
\begin{eqnarray}
\hspace{-2.5cm}\frac{M}{2\hbar}\frac{\partial_t (\delta\psi+\delta\psi^*)}{\psi_0}
+\frac{M}{\hbar}\frac{\partial_t\delta\sigma}{\sigma_0}=
-\frac{i}{4\hbar^2R^2\psi_0}L^2(\delta\psi-\delta\psi^*)
-\left(
  \frac{3\sigma_0^2}{8\hbar^2R^2}L^2+1
\right)
\delta B. \label{difeq3}
\end{eqnarray}
Here we have used that for a sphere the Laplace-Beltrami operator is proportional to the square of the angular momentum operator $L^2$ via
\begin{equation}
\displaystyle \Delta_{\mathcal{M}}=-\frac{L^2}{\hbar^2R^2}.
\end{equation}
Note that the eigenvalues of $L^2$ are given by $\hbar^2 l(l+1)$, for $l=0,1,2,...$ being the angular momentum quantum numbers and its eigenfunctions are proportional to the spherical harmonics $Y_{lm}$, with $m=0,\pm1,...,\pm l$.

The technical details on how to solve equations~(\ref{difeq1})--(\ref{difeq3}) are relegated to appendix~B. There it is shown that these equations can be straight-forwardly solved by decomposing the functions $\delta\psi$, $\delta\sigma$, $\delta B$ for all $l=0,1,2...$ in terms of $\bar{Y}_{lm}=Y_{lm}+Y_{lm}^*$ for $m=0,...,l$ and proportional to $\bar{Y}_{lm}=-i(Y_{lm}-Y_{lm}^*)$ for $m=-l,...,-1$. Here we restrict ourselves to summarize and discuss the respective results. 

For $l=0$, the collective oscillation mode frequency is given by
\begin{equation}
\Omega_0=\omega\sqrt{3+\frac{\sigma_{\rm osc}^4}{\sigma_0^4}}, \label{Omega0}
\end{equation}
which is of the order of the transversal confinement frequency $\omega$.
It coincides with the formula obtained in reference~\cite{2007}, where the mode associated to this frequency was called \emph{accordion mode}.

For $l\geq 1$, irrespective of the sign of the dimensionless interaction strength $P$ we find two branches of collective oscillation mode frequencies, a larger one and a lower one, but degenerate with respect to the magnetic quantum number $m$. The branch with larger oscillation frequencies is approximately given by
\begin{equation}
\Omega_l=\Omega_0+\frac{\omega^4}{\Omega_0^3}
   \left(
     \frac{11\sigma_0^2}{8\sigma_{\rm osc}^2}
     +\frac{7\sigma_{\rm osc}^2}{4\sigma_0^2}
     +\frac{7\sigma_{\rm osc}^6}{8\sigma_0^6}
   \right)\delta_l, \label{Osc''}
\end{equation}
while the branch with smaller oscillation frequencies reads approximately
\begin{eqnarray}
\Lambda_l=\frac{\omega^2}{\Omega_0}\Bigg[\frac{P\sigma_{\rm osc}}{2\sigma_0}\left(
      5+\frac{3\sigma_{\rm osc}^4}{\sigma_0^4}
    \right)\delta_l \label{Osc'}
\\  \hspace{1.5cm} +\frac{\omega^4}{\Omega_0^4}
 \left(
      -\frac{5\sigma_0^4}{4\sigma_{\rm osc}^4}
      +\frac{45}{4}       
      +\frac{11\sigma_{\rm osc}^4}{4\sigma_0^4}
      +\frac{7\sigma_{\rm osc}^8}{4\sigma_0^8}
      +\frac{3\sigma_{\rm osc}^{12}}{2\sigma_0^{12}}
   \right)\delta_l^2\Bigg]^{1/2}.  \nonumber
\end{eqnarray}
Here
\begin{equation}
\delta_l=\frac{\sigma_{\rm osc}^2}{R^2}l(l+1) \label{deltal}
\end{equation}
represent smallness parameters, since the width $\sigma_{\rm osc}$ is supposed to be much smaller than the radius $R$ of the sphere. As discussed at the end of section~\ref{SecEquilSphere}, we consider $\sigma_{\rm osc}^2/R^2=0.01$ to be realistic for a bubble trap in microgravity. 

Note that also our dynamical results recover the corresponding ones for an infinite plane in the limit of an infinite curvature radius, i.e. $R\rightarrow\infty$. To this end the smallness parameter~(\ref{deltal}) has just to be redefined via $\sigma_{\rm osc}^2(k_x^2 + k_y^2)$, where $k_x, k_y$ denote the components of a 2D wave vector. This means that the collective frequencies~(\ref{Osc''}) and (\ref{Osc'}) still hold but represent each a continuous spectrum above a ground frequency. The latter is given in case of the upper branch~(\ref{Osc''}) by the minimal value $\Omega_0$ given in equation~(\ref{Omega0}), whereas it vanishes for the lower branch~(\ref{Osc'}).

\begin{figure}[t]
\includegraphics[scale=0.45]{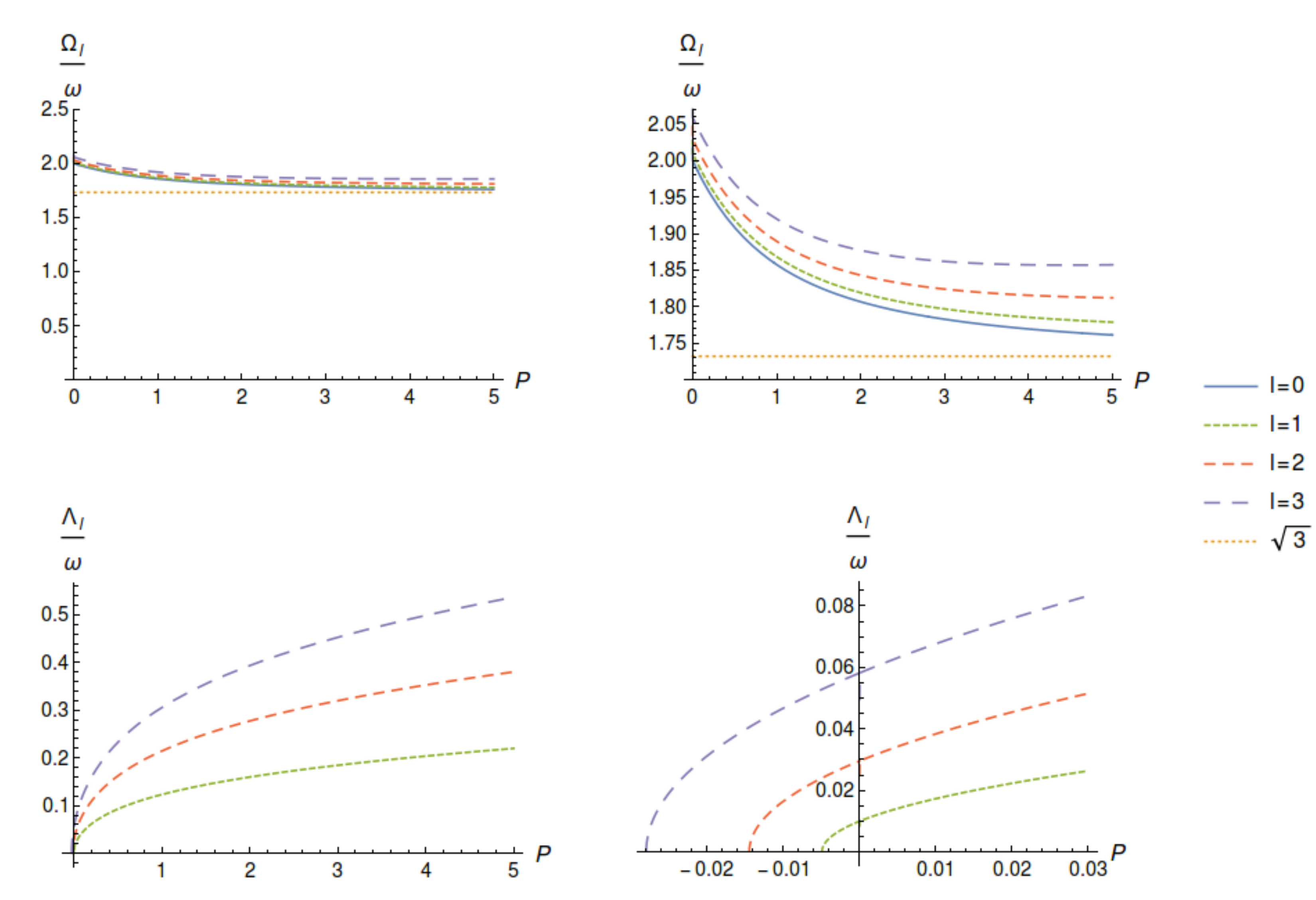}
\vspace{-10.3cm}
 
\begin{footnotesize}
(a) \hspace{6.35cm} (b)

\vspace{10.3cm}
\vspace{-5.4cm}
 
(c) \hspace{6.35cm} (d)
\end{footnotesize}
\vspace{4.4cm}
\caption{Frequencies~(\ref{Omega0})--(\ref{Osc'}) as functions of dimensionless interaction strength $P$. (a) Higher branch frequency $\Omega_l/\omega$ for $l=0,1,2,3$. (b) The same graphic as in (a), but plotted for an enlarged scale. (c) Lower branch frequency $\Lambda_l/\omega$ for $l=1,2,3$. (d) The same graphic as in (c), but plotted around $P=0$.
} \label{Freq}
\end{figure}

A plot of the frequencies $\Omega_l$ and $\Lambda_l$ as functions of the dimensionless interaction strength $P$ for the lowest values of $l$ can be found in figure~\ref{Freq}. These graphics are made for the realistic range of $P$ values, according to the parameters given at the end of section~\ref{SecEquilSphere}.

From equations~(\ref{PolynSigmaP}) and~(\ref{Omega0}) we read off that $\Omega_0/\omega$ equals to $2$ for vanishing interaction and approaches asymptotically to $\sqrt{3}$ for a large dimensionless interaction strength $P$, as can be seen in panels~\ref{Freq}(a) and (b). For $l=1,2,3$, the frequencies $\Omega_l/\omega$ have a similar behaviour as $\Omega_0/\omega$, but they turn out to be larger than $\Omega_0/\omega$. From the plots we can also see that frequencies $\Omega_l/\omega$ increase with the angular momentum quantum number $l$.

The dimensionless frequencies $\Lambda_l/\omega$ are positive for vanishing interaction, even though they are much smaller than $1$. The frequencies $\Lambda_l/\omega$ reach zero for some negative value of $P$ and monotonically increase with $P$, see figure~\ref{Freq}(c) and (d). From equation~(\ref{Osc'}) we see that these frequencies decrease for a smaller value of the parameter $\delta_l$. On the other hand, for some negative value of the dimensionless interaction strength $P$ the lower frequencies could become imaginary, such that the corresponding solution exhibit an exponential behaviour. We stress that only for quite small negative values of $P$ we still have a stable solution, as can be seen in figure~\ref{Freq}(d), and they turn out to be unstable as soon as $P$ is decreased even to relatively moderate values. Note that for $l=0$, the lower frequency is not defined due to the conservation of the particle number, as is discussed in more details in appendix B.

\section{Analysis of Modes} \label{Modes}

Within a linear stability analysis of small perturbations on a sphere, we have derived analytic expressions for two types of collective oscillation mode frequencies $\Omega_l$ and $\Lambda_l$, as well as understood their dependences on the angular momentum quantum number $l$ and on the dimensionless interaction strength $P$. Now, we analyse the respective density profiles of these oscillations on the sphere, whose calculations are relegated to appendix B. We first discuss the accordion mode, which occurs for the angular momentum quantum number $l=0$, and afterwards we analyse the modes for larger angular momentum quantum numbers $l$, and also illustrate some examples. Finally, we discuss in detail the direction of the oscillations, which can be in the confinement direction, along the surface of the sphere, or even have a mixed behaviour.

For $l=0$ only the mode with frequency~(\ref{Omega0}) appears. In this case, the temporal evolution of each component of the wave function associated with the frequency $\Omega_0$ turns out to be
\begin{eqnarray}
&\delta\psi_{00}(t)=iC_{00}\frac{\psi_0\omega}{\Omega_0}
  \left(
    \frac{P\sigma_{\rm osc}^2}{4\sigma^2_0}
    +\frac{\sigma_0}{\sigma_{\rm osc}}
  \right)
     \sin(\Omega_0 t){\bar Y}_{00}, \nonumber
\\
&\delta\sigma_{00}(t)=C_{00}\sigma_{\rm osc}\cos(\Omega_0 t){\bar Y}_{00}, \label{l0tempevol} 
\\
&\delta B_{00}(t)=C_{00}\frac{\Omega_0}{\sigma_{\rm osc}\sigma_0\omega}\sin(\Omega_0 t){\bar Y}_{00}, \nonumber
\end{eqnarray}
where $C_{00}$ is a proportionality constant defined by the intensity of the perturbation, which has to be small. Note that we have Re$\delta\psi_{00}=0$ in order to satisfy the conservation of the particle number, see appendix B for further details. 

\begin{figure}[t]
\includegraphics[scale=0.27]{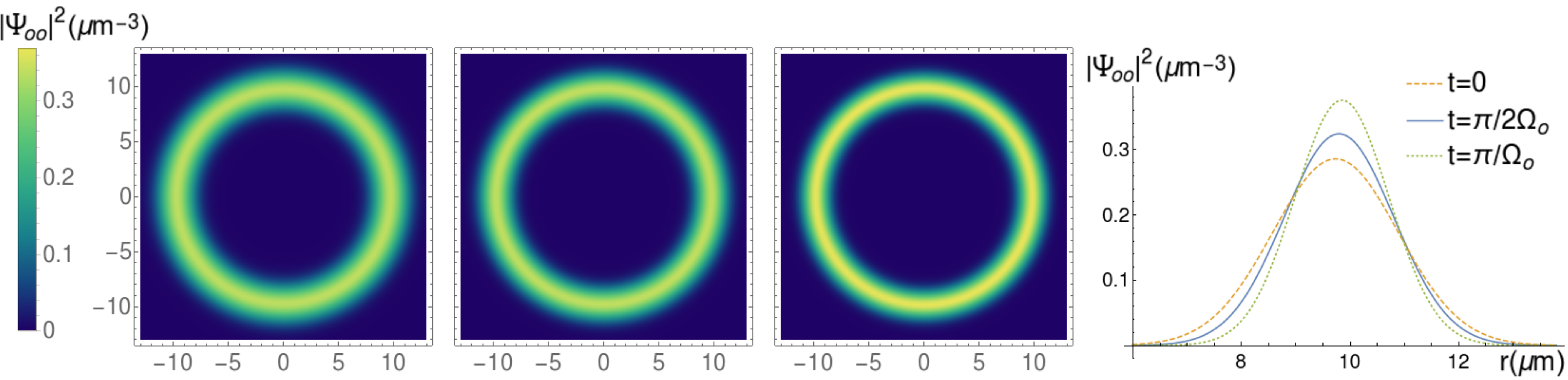}
\vspace{-4.7cm}
\begin{footnotesize}

\hspace{1.2cm} (a) \hspace{2.52cm} (b) \hspace{2.52cm} (c) \hspace{2.8cm} (d)

\end{footnotesize}
\vspace{3.7cm}
\caption{Accordion mode oscillation: density profile $|\Psi(r,\theta,\varphi,t)|^2$ at (a) $t=0$, (b) $t=\pi/2\Omega_0$ and (c) $t=\pi/\Omega_0$, in $x$--$z$ plane. (d) Radial density profiles of (a) (in orange dashed), (b) (in blue continuous), and (c) (in green dotted), for fixed $\varphi=0$ and $\theta=0$. Here $r$, $\theta$ and $\varphi$ denote the spatial variables in terms of the spherical coordinates.}
\label{BreathMode}
\end{figure}

To illustrate the accordion mode, the evolution of its density profile given by the squared norm $|\Psi_{00}(r,\theta,\varphi,t)|^2$ of the 3D wave function~(\ref{AnsatzB}) is pictured in figure~\ref{BreathMode}(a)--(c) at different times in $x$--$z$ plane, for the chosen parameters at the end of section~\ref{SecEquilSphere}. The proportionality constant is chosen to be $C_{00}=0.1$. The Gaussian width of the state in (b) coincides with the one of the  equilibrium state, while the width of the states in (a) and in (c) are, respectively, larger and smaller than the one of the equilibrium state. From the initial state in (a) which has the largest width, the system evolves to the state shown in (b) and finally reaches the state with the smallest width in (c). Then it returns to the state in (b), to the state in (a), and so on. This oscillation happens with frequency $\Omega_0$. The radial density profile of these three stages for fixed $\varphi=0$ and $\theta=0$ are plotted in figure~\ref{BreathMode}(d). The thinnest stage in green shows a higher peak and the thickest in orange shows a smaller one. This happens since the number of particles $N$ is conserved.

For the cases where $l\geq 1$, there are two types of oscillation frequencies $\Omega_l$ and $\Lambda_l$, given in equations~(\ref{Osc''}) and (\ref{Osc'}), respectively. For $\Omega_l$, the temporal evolution of the associated components of the wave function are given by
\begin{eqnarray}\label{Dyn''}
\hspace{-2.5cm} \delta\psi_{lm}(t')=C_{lm}^1\psi_0
  \Bigg\{
    -\frac{P\sigma_{\rm osc}^2\omega^2}{4\sigma_0^2\Omega_0^2}\delta_l\cos(\Omega_l t)
   \nonumber   \\ \hspace{-1cm}   +\frac{i\omega}{\Omega_0}
  \Bigg[
     \frac{\sigma_0}{\sigma_{\rm osc}}+\frac{P\sigma_{\rm osc}^2}{4\sigma^2_0}
+\frac{P\omega^4}{32\Omega_0^4}
       \left(    
          5-\frac{82\sigma_{\rm osc}^4}{\sigma^4_0}-\frac{35\sigma_{\rm osc}^8}{\sigma^8_0}
       \right)
     \delta_l
   \Bigg]\sin(\Omega_l t)
  \Bigg\}\bar{Y}_{lm}, \nonumber \\
\hspace{-2.5cm}\delta\sigma_{lm}(t')=C_{lm}^1\sigma_{\rm osc}\cos(\Omega_l t)\bar{Y}_{lm}, \ \ \ \ \  \\
\hspace{-2.5cm}\delta B_{lm}(t')=C_{lm}^1\frac{\Omega_0}{\sigma_{\rm osc}\sigma_0\omega}
  \left[
    1-\frac{P\omega^4}{\Omega^4_0}
      \left(
        \frac{7\sigma_{\rm osc}^2}{\sigma_0^2}
        +\frac{5\sigma_{\rm osc}^6}{\sigma_0^6}
      \right)\delta_l
  \right]\sin(\Omega_l t)\bar{Y}_{lm}. \nonumber
\end{eqnarray}
where $C_{lm}^1$ is a proportionality constant. For $\Lambda_l$ the solutions are
\begin{eqnarray}
\hspace{-2.5cm}\delta\psi_{lm}(t)=C_{lm}^2\psi_0
  \left(
     \frac{1}{2}\cos(\Lambda_l t)
    -i\frac{\Lambda_l}{\omega\delta_l}
  \left(
    1-\frac{P\sigma_{\rm osc}\omega^2}{4\sigma_0\Omega^2_0}\delta_l
  \right)\sin(\Lambda_l t)
  \right)\bar{Y}_{lm}, \nonumber \\
\hspace{-2.5cm}\delta\sigma_{lm}(t)=C_{lm}^2\frac{P\sigma_{\rm osc}^3\omega^2}{\sigma_0^2\Omega_0^2}\left\{1+
  \left[
    -\frac{\sigma_{\rm osc}^2\omega^2}{\sigma^2_0\Omega_0^2}
    +\frac{P\sigma_{\rm osc}\omega^4}{2\sigma_0\Omega_0^4}
      \left(
        5+\frac{3\sigma_{\rm osc}^4}{\sigma^4_0}
      \right)
  \right]
\delta_l\right\}\cos(\Lambda_l t)
\bar{Y}_{lm},
 \ \ \ \ \  \label{Dyn'} \\
\hspace{-2.5cm}\delta B_{lm}(t')=C_{lm}^2\frac{P\sigma_{\rm osc}\Lambda_l\omega}{\sigma_0^3\Omega_0^2}\left\{1+
  \left[
    -\frac{\sigma_{\rm osc}^2\omega^2}{\sigma^2_0\Omega_0^2}
    +\frac{P\sigma_{\rm osc}\omega^4}{2\sigma_0\Omega_0^4}
      \left(
        5+\frac{3\sigma_{\rm osc}^4}{\sigma^4_0}
      \right)
  \right]
\delta_l\right\}\sin(\Lambda_l  t)
\bar{Y}_{lm},  \nonumber
\end{eqnarray}
with $C_{lm}^2$ also being a proportionality constant.

\begin{figure}[t]
\includegraphics[scale=0.27]{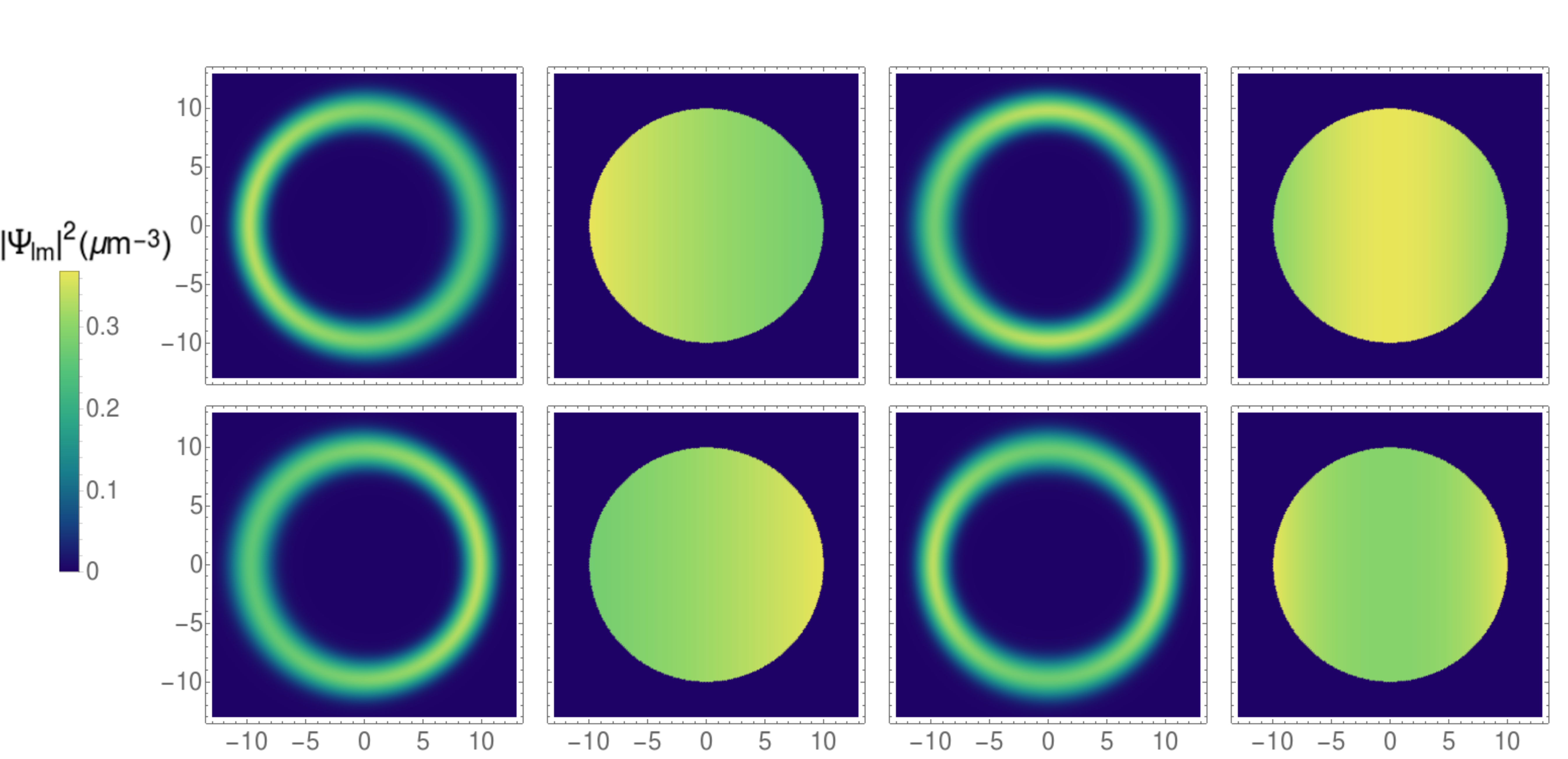}
\vspace{-7.4cm}
\begin{footnotesize}

\hspace{1.5cm} (a) \hspace{2.6cm} (b) \hspace{2.6cm} (c) \hspace{2.6cm} (d)

\end{footnotesize}
\vspace{7.1cm}

\caption{Illustration of condensate density $|\Psi(r,\theta,\varphi,t)|^2$ for $l=1$ and $l=2$, with $m=0$, at $t=0$ (top row) and $t=\pi/\Omega_l$ (bottom row). (a) Density profile for $l=1$ in $x$--$z$ plane. (b) Density on surface of the sphere for $l=1$, i.e, $|\Psi(R,\theta,\varphi,t)|^2$. (c) Profile for $l=2$ in $x$--$z$ plane. (d) Density on surface of the sphere for $l=2$.} \label{Oscl12}
\end{figure}

We illustrate the density profiles of these modes in figure~\ref{Oscl12} for $l=1$ and $l=2$, with $m=0$. Figures~\ref{Oscl12}(a) and (c) show the density profile $|\Psi(r,\theta,\varphi,t)|^2$ in the $x$--$z$ plane. Figures~\ref{Oscl12}(b) and (d) show the condensate density on the surface of the sphere, i.e., $|\Psi(R,\theta,\varphi,t)|^2$. The proportionality constants are chosen to be $C^1_{10}=C^1_{20}=0.1$. In these cases, both the width $\sigma_{lm}(t)=\sigma_0+\delta\sigma_{lm}(t)$ and the density of the 2D wave function $|\psi_{lm}(t)|^2=|\psi_0|^2+2|\psi_0|\textrm{Re}\delta\psi_{lm}(t)$ turn out to oscillate in time.

From panels~\ref{Oscl12}(a) and (c) we read off that the regions on the sphere where the density maxima are located have the minimal width. The oscillations of $\delta\sigma$ change the shape of the Gaussian, similarly to what happens in figure~\ref{BreathMode}(d). The difference is that for the accordion mode this happens in a spherically symmetric way, while for larger values of $l$ there is an angular dependence, as is illustrated in figures~\ref{Oscl12}(a) and (c). Oscillations of Re$\delta\psi$ do not change the width of the Gaussian nor its shape, and only the amplitude of the wave function is changed. For $l=0$ such oscillations do not occur, because they would change the number of particles in the system. For $l\geq 1$ these oscillations change the distribution of particles along the sphere according to the spherical harmonics, and figures~\ref{Oscl12}(b) and (d) illustrate this change of the density on the surface of the sphere.

\begin{figure}
\hspace{0.3cm}\includegraphics[scale=0.65]{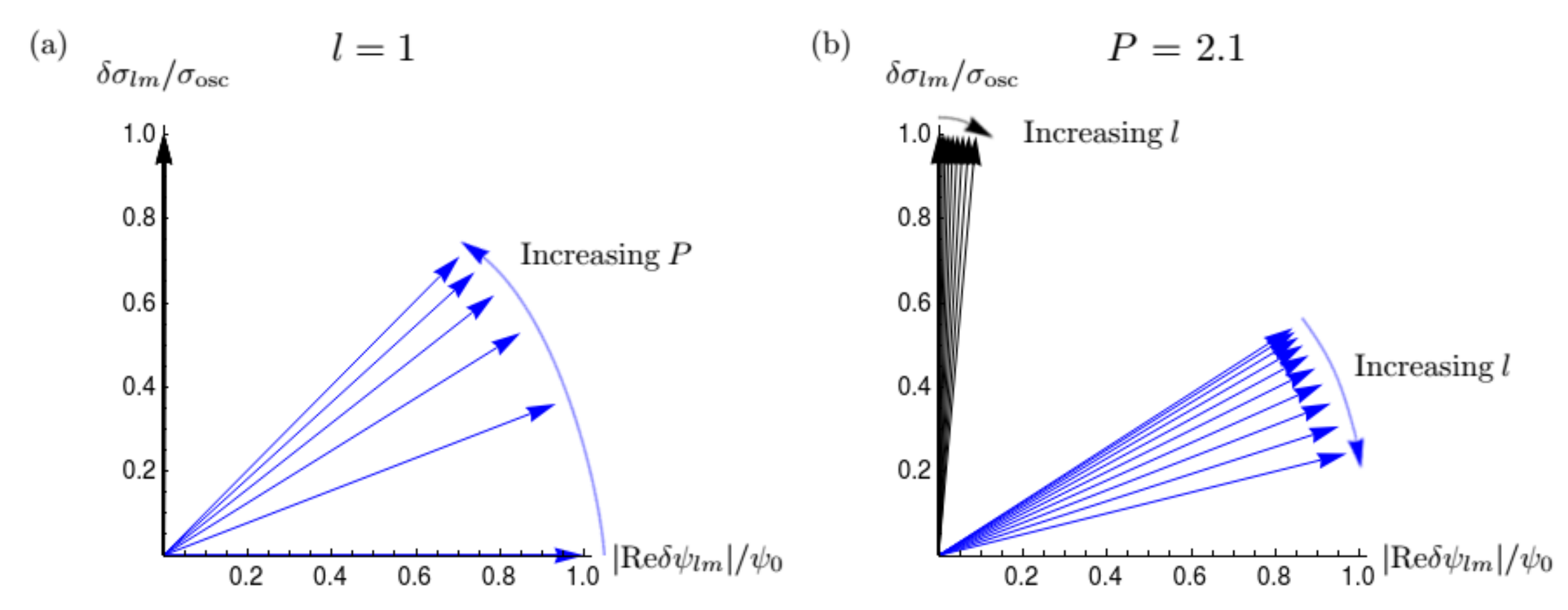}
\caption {Normalized vectors in the direction of $(|\textrm{Re}\delta\psi_{lm}|/\psi_0,\delta\sigma_{lm}/\sigma_{\rm osc})$ for various values of $l$ and $P$, for the modes corresponding to the frequencies $\Omega_l$ (in black) and $\Lambda_l$ (in blue).
(a) Fixed $l=1$ and $P=0,...,5$. Note that all black arrows practically lie on $y$ axis, and that the blue arrow corresponding to $P=0$ lies on the $x$ axis. The blue vectors increase their angle with the horizontal for increasing $P$.
(b) Fixed $P=2.1$, and $l=0,...,10$ for the larger frequencies $\Omega_l$ and $l=1,...,10$ for the lower frequencies $\Lambda_l$. Both black and blue vectors increase their angle with the vertical for increasing $l$.
\label{InitialPert}
}
\end{figure}

For the modes with frequencies $\Omega_l$ given in equation~(\ref{Dyn''}), the amplitude of $\delta\sigma_{lm}$ is much larger than the amplitude of the real part of $\delta\psi_{lm}$, since the latter one is proportional to $\delta_l$, which is small. In this sense, we say that the density oscillations predominantly occur in the direction {\it perpendicular} to the sphere. For the modes corresponding to $\Lambda_l$ given in equation~(\ref{Dyn'}), we see that the oscillations of Re$\delta\psi_{lm}$ have a fixed amplitude, while the amplitude oscillations of $\delta\sigma_{lm}$ depend on the interaction strength $P$. This dependence occurs explicitly and implicitly, since both $\sigma_0$ and $\Omega_0$ depend on $P$ according to~(\ref{PolynSigmaP}) and (\ref{Omega0}). If $P$ is zero, the amplitude oscillations of $\delta\sigma_{lm}$ vanish, so we can say that the mode is in a {\it parallel} direction to the sphere. If $P$ increases, the amplitude oscillations of $\delta\sigma_{lm}$ increase and it can even be comparable to the amplitude oscillations of Re$\delta\psi_{lm}$. In this sense, we say that the direction of oscillations turns out to have a mixed behaviour with both the parallel and perpendicular components, which we call a {\it diagonal} oscillation.

In order to illustrate the above statements, in figure~\ref{InitialPert} we plot the normalized vectors in the direction of the vectors $(|\textrm{Re}\delta\psi_{lm}|/\psi_0,\delta\sigma_{lm}/\sigma_{\rm osc})$ for various values of $l$ and $P$ in order to illustrate the difference in the contributions of $\textrm{Re}\delta\psi_{lm}$ and $\delta\sigma_{lm}$ in the modes associated to the frequencies $\Omega_l$ and $\Lambda_l$, respectively. Note that the amplitude of the collective modes have a degeneracy on the values of $m$ for $l$ fixed. Moreover, the directions of these vectors coincide with the notion given above, of oscillations being in a direction perpendicular, parallel or diagonal to the sphere.

In panel~\ref{InitialPert}(a) the angular momentum quantum number $l=1$ is fixed and we vary the dimensionless interaction strength $P$ from $0$ to $5$. We see that the amplitude oscillations for the modes with larger frequency almost do not change, while the amplitude oscillations for the modes corresponding to the smaller frequency are quite sensitive to the values of $P$. The amplitude oscillation that is parallel to the sphere for $P=0$ turns to be diagonal as $P$ increases.

In panel~\ref{InitialPert}(b) the dimensionless interaction strength $P=2.1$ is fixed and we vary the angular momentum quantum number $l$. For the modes corresponding to $\Omega_l$, $l$ is varied from $0$ to $10$, while for the modes corresponding to $\Lambda_l$, $l$ is varied from $1$ to $10$, since there is no such mode for $l=0$. We see that for both branches the vectors increase their angle with the vertical for increasing $l$, but this is more pronounced for the modes associated with the frequencies $\Lambda_l$.

From figure~\ref{InitialPert} we see that the modes corresponding to the $\Omega_l$ branch predominantly oscillate in the perpendicular direction, with a negligible dependence on $P$ and a small dependence on $l$. On the other hand, the modes from the branch corresponding to $\Lambda_l$ depend on both $P$ and $l$, but have a dominant parallel component. With this, we conclude that the oscillations with higher frequencies are in the direction of the confinement trap, while the oscillations with smaller frequencies occur along the sphere, i.e., in the not confined direction.

\section{Conclusions}

Motivated by recent experimental advances in the field, we have studied both the static and the dynamic properties of a Bose-Einstein condensate confined on the surface of a curved manifold. To this end, we provided a general formulation of the problem and derived a self-consistent set of equations for the 2D condensate wave function on the manifold and its width. In particular, we found an effective potential in the resulting 2D Gross-Pitaevskii equation, which depends on the metric of the manifold in a non-trivial way but vanishes for a sphere. For the latter special case we determined in equilibrium how both the width of the condensate and its chemical potential increase with the repulsive interaction strength.
Moreover, we found via a linear stability analysis two branches of collective excitations, with distinctly different frequencies. The larger branch frequencies turned out to be of the order of the harmonic confinement frequency and, thus, correspond to oscillations predominantly in the direction of the confinement, i.e, in the perpendicular direction to the sphere. The lower branch frequencies are much smaller and represent oscillations predominantly on the sphere. Our results represent concrete predictions, which presumably could be confirmed in upcoming
bubble trap experiments in the NASA Cold Atom Laboratory at the International Space Station~\cite{Space4}. But in order to become experimentally more realistic it would be necessary to deal with a Bose gas confined on an ellipsoid as it represents a better approximation to the bubble trap, than the sphere \cite{GarHelReview2,Space3}.

Note that the collective modes analysed in this paper differ from the ones identified in reference~\cite{Lima}. This is due to the fact that our ansatz~(\ref{AnsatzB}), (\ref{raizgt}) for the wave function neglects the possibility that the condensate levitates below or above the minimum of the confinement potential.
 
\section{Acknowledgments}

The authors thank Antun Bala\v{z} for valuable suggestions at all the stages of the project, as well as Arnol Garcia, Aristeu Lima, H\'el\`ene Perrin, Milan Radonji\'c, Luca Salasnich, Enrico Stein, and Andrea Tononi for useful discussions.

N.S.M., F.E.A.S., and A.P. thank the binational project between CAPES (Coordena{\c c}\~ao de Aperfei{\c c}oamento de Pessoal de N\'ivel Superior, Improvement Coordination of Higher Level Personnel) and DAAD (Deutscher Akademischer Austauschdienst, German Academic Exchange Service) via Probral (Programa Brasil-Alemanha, Brazil-Germany Program) No. 488/2018 Grant No. 88881.143936/2017-01, through which N.S.M. was supported by a scholarship financed by CAPES -- Brazilian Federal Agency for Support and Evaluation of Graduate Education within the Ministry of Education of Brazil.
F.E.A.S. thanks CNPq (Conselho Nacional de Desenvolvimento Científico e Tecnológico, National Council for Scientific and Technological Development) for support through Bolsa de produtividade em Pesquisa Grant No. 305586/2017-3. F.E.A.S. and V.S.B. thank CEPID / CEPOF-FAPESP program (Grant number 2013 / 07276-1). A.P. thanks the Deutsche Forschungsgemeinschaft (DFG, German Research Foundation) via the Collaborative Research Centers SFB/TR185 (project No. 277625399). 

\

\appendix

\section{The Fundamental Forms of a Manifold and the Gaussian Normal Coordinate System}

There are two fundamental forms~\cite{Manfredo} which are used to characterize a 2D manifold~$\mathcal{M}$ embedded in the 3D space. The first fundamental form is simply the metric $g_{ij}$ of the manifold~$\mathcal{M}$, as discussed in section~\ref{SecPrel}. This is a natural instrument constructed to treat lengths of curves, areas of regions as well as other metric quantities, and its expression is given by
\begin{equation}
g_{ij}={\bf v}_i\cdot{\bf v}_j, \label{FirstForm}
\end{equation}
where ${\bf v}_1$ and ${\bf v}_2$ are defined in equation~(\ref{TangentV}). Note that this is a symmetric form.

The second fundamental form is constructed to deal with changes of the normal vector along a path on the surface, providing information about its curvature, and is defined by
\begin{equation}
s_{ij}={\bf v}_i\cdot \partial_j {\bf v}_0,  \label{SecondForm}
\end{equation}
where ${\bf v}_0$ is the normal vector given by equation~(\ref{DefNormVector}). Note that $s_{ij}$ is also a symmetric form, meaning that $s_{ij}=s_{ji}$. In order to see that, we start from the orthogonality relations ${\bf v}_i\cdot {\bf v}_0=0$ and ${\bf v}_j\cdot {\bf v}_0=0$, i.e., $\partial_i{\bf p}\cdot{\bf v}_0=0$ and $\partial_j{\bf p}\cdot{\bf v}_0=0$, for $i,j=1,2$. Differentiating the first expression with respect to $x_j$ and the second with respect to $x_i$ we obtain
\begin{eqnarray}
\partial_{ij}{\bf p}\cdot{\bf v}_0+{\bf v}_i\cdot \partial_j{\bf v}_0=0, \label{A3} \\
\partial_{ji}{\bf p}\cdot{\bf v}_0+{\bf v}_j\cdot \partial_i{\bf v}_0=0. \label{A4}
\end{eqnarray}
From the symmetry of the second derivatives and (\ref{A3}), (\ref{A4}), we then conclude that ${\bf v}_i\cdot \partial_j{\bf v}_0={\bf v}_j\cdot \partial_i{\bf v}_0$. Thus, equation~(\ref{SecondForm}) can be rewritten as $s_{ij}=\frac{1}{2}({\bf v}_i\cdot \partial_j {\bf v}_0+{\bf v}_j\cdot \partial_i {\bf v}_0)$, proving the above statement that $s_{ij}$ is symmetric.

Now, consider the operator $s\cdot g^ {-1}$, which is known in the literature as the Gauss map. The directions of its eigenvectors ${\bf e}_1$ and ${\bf e}_2$ are called \emph{principal directions}, and these directions correspond to the minimal and maximal curvatures. The matrix representing $s\cdot g^ {-1}$ is given in this basis by
\begin{equation}
s\cdot g^ {-1}=\left(
\begin{array}{cc}
\kappa_1 & 0 \\ 
0 & \kappa_2
\end{array} 
\right), \label{MatSec}
\end{equation}
where $\kappa_1$ and $\kappa_2$ denote the curvatures of the manifold in the direction of ${\bf e}_1$ and ${\bf e}_2$, respectively. The {\it mean} and the {\it Gaussian curvatures} of this manifold are defined by
\begin{equation}
H=\frac{\textrm{Tr}(s\cdot g^ {-1})}{2}=\frac{\kappa_1+\kappa_2}{2} \ , \ \ \ \ \ 
K=\det (s\cdot g^ {-1})=\kappa_1\kappa_2.
\end{equation}
Furthermore, there is a third fundamental form~\cite{III} defined as
\begin{equation}
h_{ij}= \partial_i {\bf v}_0\cdot \partial_j {\bf v}_0. \label{ThirdForm}
\end{equation}
It can be shown that this form turns out to be a combination of the previous two, via the relation~\cite{III}
\begin{equation}
h_{ij}=-Kg_{ij}+2Hs_{ij},
\end{equation}
and the matrix which represents $h\cdot g^ {-1}$ in the principal directions is given by
\begin{equation}
h\cdot g^ {-1}=\left(
\begin{array}{cc}
\kappa_1^2 & 0 \\ 
0 & \kappa_2^ 2
\end{array} 
\right).   \label{MatThird}
\end{equation}
Now, let us consider a manifold $\mathcal{M}(x^0)$ parallel to the manifold $\mathcal{M}$, as explained in section~\ref{SecPrel}. The metric, i.e., the first fundamental form of this manifold $\mathcal{M}(x^0)$, for $|x^0|< R/2$, is defined by
\begin{equation}
g_{ij}(x^0,x^1,x^2)=\partial_i {\bf q}\cdot\partial_j {\bf q}
\end{equation}
and from expression~(\ref{GaussNorm}) we conclude
\begin{equation}
g_{ij}(x^0,x^1,x^2)=({\bf v}_i+x^0\partial_i {\bf v}_0)\cdot({\bf v}_j+x^0\partial_j {\bf v}_0).
\end{equation}
Combining this equation with the definitions~(\ref{FirstForm}), (\ref{SecondForm}), and (\ref{ThirdForm}), and taking into account that the fundamental forms are symmetric, we read off that this metric of the manifold $\mathcal{M}(x^0)$ can be expressed in terms of the three fundamental forms of the manifold $\mathcal{M}$ as follows:
\begin{equation}
g_{ij}(x^0,x^1,x^2)=g_{ij}+2x^0s_{ij}+(x^0)^2h_{ij}.
\end{equation}
From the corresponding matrix forms~(\ref{MatSec}) and (\ref{MatThird}), it is then possible to deduce
\begin{equation}
\hspace{-1.5cm}\det g(x^0)=\det g \left[
                     1+4x^0 H+(x^0)^2(4H^2+2K)
                     +4(x^0)^3HK+(x^0)^4K^2
                   \right].
\end{equation}

Thus, we obtain
\begin{equation}
\det g(x^0)\leq \det g \left[
                     1+\frac{4x^0}{R}
                     +\frac{6(x^0)^2}{R^2}
                     +\frac{4(x^0)^3}{R^3}
                     +\frac{(x^0)^4}{R^4}
                   \right],
\end{equation}
where $R$ is defined as the minimum mean radius over all points ${\bf p}$ belonging to $\mathcal{M}$ according to equation~(\ref{R}). The square root of this formula can be expanded in a Taylor series, yielding equation~(\ref{Taylor}) from the main text.

\section{Determining the Collective Modes}

In this section we present the detailed calculations for the respective results of sections~\ref{ColModes} and \ref{Modes}. For the sake of simplicity, we will use a dimensionless form for our equations. To this end, we define the following dimensionless variables
\begin{equation}
\sigma'=\frac{\sigma}{\sigma_{\rm osc}} \ , \ \ \ \ B'=\sigma_{\rm osc}^2B \ , \ \ \ \ 
\psi'=\frac{\psi}{\psi_0} \ , \ \ \ \ t'=\omega t,
\end{equation}
and the dimensionless derivative operators
\begin{equation}
\partial_{t'}=\frac{\partial_t}{\omega} \ , \ \ \ \ L'^2=\frac{L^2}{\hbar^2}.
\end{equation}
With that, the dimensionless form of the linearized equations~(\ref{difeq1})--(\ref{difeq3}) reads
\begin{eqnarray}
\label{LinearPsi}
\hspace{-2.5cm} i\partial_{t'}\delta\psi'
+\frac{\sigma_0'^2}{4}\partial_{t'}\delta B'
=&
\frac{\sigma_{\rm osc}^2}{2R^2}L'^2\delta\psi'
+\left(
    \sigma_0'^2-\frac{1}{\sigma_0'^2}
 \right)
(\delta\psi'+\delta\psi'^*)
\\
\nonumber
&
+\frac{1}{2}
 \left(
    \frac{1}{\sigma_0'^3}-\sigma_0'
 \right)
\delta\sigma' 
-\frac{i\sigma_0'^2}{8}\frac{\sigma_{\rm osc}^2}{R^2}L'^2\delta B',
\end{eqnarray}
\begin{equation}
\hspace{-2.5cm}\partial_{t'}\delta B'=
\left(
  \frac{1}{\sigma_0'^4}-1
\right)
(\delta\psi'+\delta\psi'^*)
+
\left(
  3+\frac{1}{\sigma_0'^4}
\right)\frac{\delta\sigma'}{\sigma_0'}
+\frac{1}{\sigma_0'^3}\frac{\sigma_{\rm osc}^2}{R^2}L'^2\delta\sigma',
 \label{LinearSigma}
\end{equation}
\begin{eqnarray}
\hspace{-2.5cm}\frac{1}{2}\partial_{t'}(\delta\psi'+\delta\psi'^*)
+\frac{\partial_{t'}\delta\sigma'}{\sigma_0'}
=-\frac{i}{4}\frac{\sigma_{\rm osc}^2}{R^2}L'^2(\delta\psi'-\delta\psi'^*)
-\delta B'
-\frac{3\sigma_0'^2}{8}\frac{\sigma_{\rm osc}^2}{R^2}L'^2\delta B'. \label{LinearB}
\end{eqnarray}
The solution of these coupled equations can be written in terms of the decompositions
\begin{equation}
\hspace{-0.5cm}\delta\psi'=\sum_{l=0}^\infty\sum_{m=-l}^l\delta\psi'_{lm} 
\ , \ \ \ \ 
\delta\sigma'=\sum_{l=0}^\infty\sum_{m=-l}^l\delta\sigma'_{lm}
 \ , \ \ \ \ 
\delta B'=\sum_{l=0}^\infty\sum_{m=-l}^l\delta B'_{lm} .
\label{DecompPsi}
\end{equation}
At this point we could have chosen the fundamental solutions $\delta\psi'_{lm}$, $\delta\sigma'_{lm}$ and $\delta B'_{lm}$ to be proportional to $Y_{lm}$. But if we had done that, the respective expansion coefficients would be coupled. Instead, we define for all $l=0,1,2,..$ the function $\bar{Y}_{lm}=Y_{lm}+Y^*_{lm}$ for $m=0,...,l$, and $\bar{Y}_{lm}=-i(Y_{lm}-Y^*_{lm})$ for $m=-1,...,-l$ and choose the following fundamental solutions
\begin{equation}
\hspace{-1.5cm}\delta\psi'_{lm}=\frac{k_{lm}(t')+ih_{lm}(t')}{2}\bar{Y}_{lm} \ , \ \ \ \ \ \ \ \delta\sigma'_{lm}=s_{lm}(t')\bar{Y}_{lm} \ , \ \ \ \ \ \ \ \delta B'_{lm}=b_{lm}(t')\bar{Y}_{lm} \ , \ \  \label{lmpm}
\end{equation}
where the coefficients $k_{lm}$, $h_{lm}$, $s_{lm}$ and $b_{lm}$ are real functions of $t'$ for all $l=0,1,2,...$ and $m=0,\pm1,...\pm l$. With this, it is possible to decouple the equations of different indices $l$ and $m$, since $\bar{Y}_{lm}^*=\bar{Y}_{lm}$. Thus, the following equations turn out to be much simpler than they would have been with the standard decomposition. 

Now we insert the ansatz~(\ref{lmpm}) into equations~(\ref{LinearPsi})--(\ref{LinearB}), by taking into account that $L'^2Y_{lm}=l(l+1)Y_{lm}$. With this, we obtain that the real part of equation~(\ref{LinearPsi}) is given by
\begin{equation}
-\frac{\partial_{t'}h_{lm}}{2}+\frac{\sigma_0'^2}{4}\partial_{t'}b_{lm}=
  \left(
    \sigma_0'^2
    -\frac{1}{\sigma_0'^2}
    +\frac{\delta_l}{4}
  \right)k_{lm}
+\frac{1}{2}\left(
              \frac{1}{\sigma_0'^3}-\sigma_0'
            \right)s_{lm}, \label{B8}
\end{equation}
while its imaginary part becomes
\begin{equation}
\frac{\partial_{t'}k_{lm}}{2}=
\frac{\delta_l}{4}h_{lm}
-\frac{\sigma_0'^2}{8}\delta_l b_{lm}.
\end{equation}
Correspondingly, equation~(\ref{LinearSigma}) reduces to
\begin{equation}
\partial_{t'}b_{lm}=
 \left(
    \frac{1}{\sigma_0'^4}-1
 \right)k_{lm}
+\left(
    \frac{1}{\sigma_0'^5}
    +\frac{3}{\sigma_0'}
    +\frac{\delta_l}{\sigma_0'^3}
\right)s_{lm},
\end{equation}
whereas equation~(\ref{LinearB}) becomes
\begin{equation}
\frac{\partial_{t'}k_{lm}}{2}+\frac{\partial_{t'}s_{lm}}{\sigma_0'}=
\frac{\delta_l}{4}h_{lm}
-\left(
   1+\frac{3\sigma_0'^2}{8}\delta_l
 \right)b_{lm}. \label{B11}
\end{equation}
Rearranging these equations they turn out to be of the form of a system of linear differential equations:
\begin{equation}
\partial_{t'}X_{lm}(t')=Q_{l}X_{lm}(t'), \label{LinearSystem}
\end{equation}
where we introduce the vector
\begin{equation}
X_{lm}(t')^T=(k_{lm}(t'),h_{lm}(t'),s_{lm}(t'),b_{lm}(t')), \label{B13} 
\end{equation}
and the matrix $Q_l=Q_0+\delta Q_l$ with
\begin{equation}
Q_{0}=\left(
\begin{array}{cccc}
0 &   0  & 0 &  0 
\\ 
\\
\displaystyle -\frac{5\sigma_0'^2}{2}+\frac{5}{2\sigma_0'^2} & 0 & \displaystyle -\frac{1}{2\sigma_0'^3}+\frac{5\sigma_0'}{2} & 0 
\\ 
\\
0 & \  0  \  & 0 & \displaystyle   -\sigma_0'
\\ 
\\
\displaystyle \frac{1}{\sigma_0'^4}-1 & 0 & \displaystyle \frac{1}{\sigma_0'^5}+ \frac{3}{\sigma_0'} &  0
\end{array}\right)
\end{equation}
and
\begin{equation}
\delta Q_{l}=\left(
\begin{array}{cccc}
0 & \displaystyle \frac{\delta_l}{2}  & 0 & \displaystyle -\frac{\sigma_0'^2}{4}\delta_l
\\ 
\\
\displaystyle -\frac{\delta_l}{2} & 0 & \displaystyle\frac{\delta_l}{2\sigma_0'} & 0 
\\ 
\\
\ \ \  \ 0 \ \  \ \  & \ \ \ \ 0 \ \ \ \    & \ \ \ \ \ 0 \ \ \ \ \ & \displaystyle -\frac{\sigma_0'^3}{4}\delta_l
\\ 
\\
0 & 0 & \displaystyle \frac{\delta_l}{\sigma_0'^3} & \ \ \ \ \ 0 \ \ \ \ \
\end{array}\right).
\end{equation}
The general solution of this system is given by $e^{Q_lt'}X_{lm}(0)$, where $X_{lm}(0)$ denotes the initial condition. 

The problem of computing the exponential matrix $e^{Q_lt'}$ applied to some vector is standard when the matrix $Q_l$ has four different eigenvectors. Let $X_i$, for $i=1,2,3,4$, be the eigenvectors associated to the eigenvalues $\lambda_i$, and $X_{lm}(0)$ to be written in the eigenbasis, i.e., $X_{lm}(0)=\sum_{i=1}^4v_iX_i$. Then the solution of (\ref{LinearSystem}) with initial condition $X_{lm}(0)$ is given by $X_{lm}(t')=\sum_{i=1}^4v_ie^{\lambda_it'}X_i$. This procedure is going to be used for solving equation~(\ref{LinearSystem}) when $l\geq 1$ in appendix B.2 and when $l=0$ and $P=0$ in the next subsection.

But when $l=0$ and $P\neq 0$, it turns out that the matrix $Q_0$ has only three eigenvectors, meaning that the eigenbasis is incomplete. In this case one has to consider generalized eigenvectors, as is shown in detail in the following subsection.

\subsection{Solving Differential Equation for $l=0$}

To find the solution of~(\ref{LinearSystem}) when $l=0$, we have to compute $e^{Q_0t'}X_{00}(0)$. The eigenvalues of $Q_0$ are $\lambda_0^{1\pm}=\pm i\Omega'_0$ with multiplicity one and $\lambda_0^2=0$ with multiplicity two. Here, $\Omega'_0=\Omega_0/\omega$ and $\Omega_0$ is given in equation~(\ref{Omega0}) of the main text. We discuss now separately the cases $P=0$ and $P\neq 0$.

When $P=0$ the respective eigenvectors are
\begin{equation}
X_{0}^{1\pm}=\left(\begin{array}{cccc}
 0  \\ 
 \mp i  \\
 1  \\ 
 \mp 2i 
\end{array}\right),\ \ \ \ \
X_{0}^{2+}=\left(\begin{array}{cccc}
 0  \\ 
 1  \\ 
 0  \\ 
 0 
\end{array}\right), \ \ \ \ \
X_{0}^{2-}=\left(\begin{array}{cccc}
 1  \\ 
 0  \\ 
 0  \\
 0 
\end{array}\right),  \label{B16}
\end{equation}
where $X_0^{1\pm}$ are associated to the eigenvalues $\pm i\Omega'_0$ and $X_0^{2\pm}$ correspond to the eigenvalue $0$. Then the initial condition is decomposed according to
\begin{equation}
X_{00}(0)=v_1X_0^{1+}+v_2X_0^{1-}+v_3X_0^{2+}+v_4X_0^{2-}, \label{**}
\end{equation}
where $v_i$, for $i=1,2,3,4$ are constants, so the exponential matrix $X_{00}(t')=e^{Q_0t'}X_{00}(0)$ yields the solution
\begin{equation}
X_{00}(t')=v_1e^{i\Omega'_0t'}X_0^{1+}+v_2e^{-i\Omega'_0t'}X_0^{1-}+v_3X_0^{2+}+v_4X_0^{2-}. \label{solutionl0P0}
\end{equation}
Let us consider now $P\neq 0$. In this case it turns out that the matrix $Q_0$ has only three eigenvectors. In order to have a basis of the four-dimensional vectorial space, we choose the vectors
\begin{equation}
\hspace{-2.5cm} X_{0}^{1\pm}=\left(\begin{array}{cccc}
0 \\ \\
\displaystyle\mp\frac{i}{\Omega'_0}
  \left(
    \frac{P}{2\sigma'^2_0}
    +2\sigma'_0
  \right)
 \\ \\
1 \\ \\
\displaystyle\mp\frac{i\Omega'_0}{\sigma'_0}
\end{array}\right), \
X_{0}^{2+}=\left(\begin{array}{cccc}
\\  0  \\ \\
 1  \\ \\
 0  \\ \\
 0  \\ \
\end{array}\right), \ 
\\ \\
\bar{X}_{0}^{2}=-\frac{\sigma'^2_0}{P
  \left(
    5+\frac{3}{\sigma'^4_0}
  \right)}
\left(\begin{array}{cccc}
 \displaystyle\frac{\Omega'^2_0}{\sigma'_0} \\ \\
 0  \\ \\
 \displaystyle\frac{P}{\sigma'^3_0}  \\ \\
 0 
\end{array}\right), \label{GeneralEigenbasis}
\end{equation}
where $X_{0}^{1\pm}$ are the eigenvectors associated to the eigenvalues $\pm i\Omega'_0$,  $X_{0}^{2+}$ is the eigenvector associated to the eigenvalue $0$ and $\bar{X}_{0}^{2}$ is a generalized eigenvector with the property $Q_0\bar{X}_0^2=X_0^{2+}$. Remember that $\sigma'_0=1$ and $\Omega'_0=2$ when $P=0$, meaning that the vectors $X_0^{1\pm}$ of equation~(\ref{GeneralEigenbasis}) reduce to the vectors $X_0^{1\pm}$ of equation~(\ref{B16}), when $P$ tends to $0$. Because of this relation, we treat formally the two different regimes, $P\neq 0$ and $P=0$, from now on as the same case when we study the solutions associated to the vectors $X_{0}^{1\pm}$.

The Jordan normal form of the matrix $Q_0$ in the basis~(\ref{GeneralEigenbasis}) is represented by the matrix
\begin{equation}
Q^\mathcal{J}_0=\left(\begin{array}{cccc}
\  i\Omega'_0  &       0       & \ 0   \  \  & 0  \ \  \\ 
0             &  -i\Omega'_0   & \ 0   \  \  & 0  \ \  \\ 
0             &      0        & \ 0   \  \  & 1  \ \  \\ 
0             &      0        & \ 0   \  \  & 0  \ \ 
\end{array}\right),
\end{equation}
while $Q_0=TQ^\mathcal{J}_0T^{-1}$, where the matrix $T=(X_{0}^{1+}$ \ $X_{0}^{1-}$ \ $X_{0}^{2+}$ \ $\bar{X}_0^{2})$ consists of the vectors~(\ref{GeneralEigenbasis}) as column vectors. With that the exponential of $Q_0t'$ is given by
\begin{equation}
\ \ e^{Q_{0}t'}=T\left(\begin{array}{cccc}
\  e^{i\Omega'_0t'} &       0                  & \ 0   \  \  & 0   \ \  \\ 
0                  &  \ \ e^{-i\Omega'_0 t'}   & \ 0   \  \  & 0   \ \  \\ 
0                  &      0                   & \ 1   \  \  & t'  \ \  \\ 
0                  &      0                   & \ 0   \  \  & 1   \ \ 
\end{array}\right)T^{-1}.
\end{equation}
With this, for an initial condition given by the vector 
\begin{equation}
X_{00}(0)=v_1X_0^{1+}+v_2X_0^{1-}+v_3X_0^{2+}+v_4\bar{X}_0^{2}, \label{*}
\end{equation}
the solution of~(\ref{LinearSystem}) reads
\begin{equation}
X_{00}(t')=v_1e^{i\Omega_0t'}X_0^{1+}+v_2e^{-i\Omega_0t'}X_0^{1-}+v_3X_0^{2+}+v_4(\bar{X}_0^{2}+t'X_0^{2+}). \label{solutionl0P}
\end{equation}
Thus, there is a secular term in the last term of the right-hand side, which leads to linear growth in time. In the following we show that this solution is eliminated due to the conservation of the number of particles, which is described by
\begin{equation}
N=\int |\psi_0|^2 dA=\int |\psi(t)|^2 dA. \label{Npsiot}
\end{equation}
The last integral in (\ref{Npsiot}) can be computed with the decomposition~(\ref{smallpsidimension})--(\ref{smallB}) by considering only the first order of the perturbation:
\begin{eqnarray}
\int |\psi(t)|^2 dA=\int [ \ |\psi_0|^2+|\psi_0|(\delta\psi'+\delta\psi'^*)]dA. \label{Npsit}
\end{eqnarray}
Comparing equations~(\ref{Npsiot}) and (\ref{Npsit}), we then conclude that
\begin{equation}
\int(\delta\psi'+\delta\psi'^*)dA=0.
\end{equation}
Using the decomposition~(\ref{DecompPsi}) for $\delta\psi'$ we obtain at first
\begin{equation}
4\pi R^2(\delta\psi_{00}'+\delta\psi_{00}'^*)+\sum_{l=1}^\infty\sum_{m=-l}^l\int(\delta\psi'_{lm}+\delta\psi'^*_{lm})dA=0. \label{EqNorm}
\end{equation}
Due to the orthogonality relations of the spherical harmonics, for all $l\geq 1$ and $m=-l,...,l$, we have that
\begin{equation}
\int \delta\psi'_{lm} dA=\int \delta\psi'^*_{lm} dA=0.
\end{equation}
With this, equation~(\ref{EqNorm}) reduces to
\begin{equation}
\delta\psi_{00}'+\delta\psi_{00}'^*=0 ,
\end{equation}
which means that $\delta\psi_{00}(t')$ is a purely imaginary number. Thus, from equation~(\ref{lmpm}) we conclude that $k_{00}(t')=0$. Due to~(\ref{B13}), (\ref{B16}), (\ref{**}), (\ref{GeneralEigenbasis}) and (\ref{*}), this fixes the coefficient $v_4=0$ in solutions (\ref{solutionl0P0}) and (\ref{solutionl0P}) for $P=0$ and $P\neq 0$, respectively. This is important for the case of equation~(\ref{solutionl0P}), because it removes the secular term.

Moreover, note that the solutions in (\ref{solutionl0P0}) and (\ref{solutionl0P}) associated to the eigenvector $X^{2+}_0$ are temporally constants. With this, we find that only the eigenvectors $X_{0}^{1\pm}$ describe oscillating solutions in time. However, we can see that the vectors $X_0^{1\pm}$ are complex, so the physical meaning is not evident here. On the other hand, we remark that the first and third entries of both vectors in (\ref{B16}) and (\ref{GeneralEigenbasis}) are real numbers, while the second and fourth entries are purely imaginary numbers. Then the linear combinations
\begin{equation}
\frac{X_{0}^{1+}+X_{0}^{1-}}{2}
\ \ \ \textrm{and} \ \ \ \ 
i\frac{(X_{0}^{1+}-X_{0}^{1-})}{2}
\end{equation}
are real vectors. With this we obtain
\begin{eqnarray}
\hspace{-2.5cm} X_{00}(t')=e^{Q_0t'}\frac{X_{0}^{1+}+X_{0}^{1-}}{2}&=\frac{e^{\lambda_0^{1+}t'}X_{0}^{1+}+e^{-\lambda_0^{1+}t'}X_{0}^{1-}}{2} \nonumber
\\
&=\cos(\Omega_0't')\frac{X_{0}^{1+}+X_{0}^{1-}}{2}+i\sin(\Omega_0't')\frac{(X_{0}^{1+}-X_{0}^{1-})}{2}, \label{B31}
\end{eqnarray}
which is a real solution. Analogously, we could compute the solution $e^{Q_0t'}i\frac{X_{0}^{2+}-X_{0}^{2-}}{2}$, but it turns out to lead to the same dynamics as above apart from a phase. So, we will consider only the solution~(\ref{B31}) which can be written as
\begin{equation}
\hspace{-2cm} X_{00}(t')^T=\left[ \ 0 \ ,\frac{C_{00}}{\Omega'_0}
  \left(
    \frac{P}{2\sigma'^2_0}
    +2\sigma'_0
  \right)
     \sin\left(\Omega'_0 t'\right),C_{00}\cos\left(\Omega'_0 t'\right),\frac{C_{00}\Omega'_0}{\sigma_0'}\sin\left(\Omega'_0 t'\right)\right]
\end{equation}
with the initial condition
\begin{equation}
X_{0}(0)^T=(0,0,C_{00},0),
\end{equation}
where $C_{00}$ is a proportionality constant. This solution means that, when one performs at $t=0$ a small perturbation of the ground state given by
\begin{equation}
\delta\sigma'_{00}(0)=C_{00}{\bar Y}_{00},
\end{equation}
then the temporal evolution of this perturbation is given by
\begin{eqnarray}
\delta\psi'_{00}(t')&=i\frac{C_{00}}{\Omega'_0}
  \left(
    \frac{P}{4\sigma'^2_0}
    +\sigma'_0
  \right)
     \sin(\Omega'_0 t'){\bar Y}_{00}, \nonumber
\\ \ \nonumber \\
\delta\sigma'_{00}(t')&=C_{00}\cos(\Omega'_0 t'){\bar Y}_{00},
\\ \ \nonumber \\
\delta B'_{00}(t')&=\frac{C_{00}\Omega'_0}{\sigma'_0}\sin(\Omega'_0 t'){\bar Y}_{00}.\nonumber
\end{eqnarray}
These equations are the dimensionless form of equations~(\ref{l0tempevol}) in the main text.

\subsection{Solving Differential Equations for $l\geq 1$}

In this section we solve equation~(\ref{LinearSystem}) for $l\geq 1$. The solution of this system is given by
\begin{equation}
X_{lm}(t')=e^{Q_lt'}X_{lm}(0), \label{Xlmt}
\end{equation}
where $X_{lm}(0)$ denotes the initial state. In order to evaluate the matrix exponential $e^{Q_lt'}$ we have  to determine at first the eigenvalues of the matrix $Q_l$. They turn out to be the roots of the polynomial
\begin{equation}
\lambda^4+(\Omega_0'^2+a_l)\lambda^2+c_l=0, \label{Poli}
\end{equation}
where we consider for the coefficients
\begin{equation}
a_l=\left(
      \frac{7\sigma_0'^2}{4}+\frac{1}{4\sigma_0'^2}
    \right)\delta_l
    +\frac{\delta_l^2}{2}
\end{equation}
and
\begin{equation}
c_l=\frac{P}{2\sigma_0'}\left(
      5+\frac{3}{\sigma_0'^4}
    \right)\delta_l
  +\left(
      \frac{3}{2}+\frac{5\sigma_0'^4}{8}-\frac{9}{8\sigma_0'^4}
   \right)\delta_l^2.
\end{equation}
both first and second orders of the smallness parameters $\delta_l$ defined in equation~(\ref{deltal}), even though for the final result of the frequencies we restrict ourselves to the first order corrections.

The roots of equation~(\ref{Poli}) are given by
\begin{equation}
\lambda_l=\pm\sqrt{\frac{-\Omega_0'^2-a_l\pm\sqrt{(\Omega_0'^2+a_l)^2-4c_l}}{2}},
\end{equation}
which can be expanded by taking into account only first and second orders of $a_l$ and $c_l$, yielding
\begin{equation}
\lambda_l^{1\pm}=\pm i\Omega_0'\left(
                         1+\frac{a_l}{2\Omega_0'^2}-\frac{c_l}{2\Omega_0'^4}                       
                       \right), \ \ \ 
\lambda_l^{2\pm}=\pm i\sqrt{\frac{c_l}{\Omega_0'^2}+\frac{c_l^2}{\Omega_0'^6}-\frac{a_lc_l}{\Omega_0'^4}}.
\end{equation}
The roots $\lambda_l^{1\pm}$ are always imaginary, so they correspond to oscillating solutions. Instead, the roots $\lambda_l^{2\pm}$ are imaginary for positive values of the radicand, leading also to oscillating solutions, but they are real when the radicand is negative, leading to exponentially increasing and decreasing solutions. The radicand is positive for positive and small negative values $P$, but it turns out to be negative for most negative values of $P$. Thus, we can conclude that the system is stable for positive interactions and small enough negative interaction strengths, while it becomes unstable for most negative interaction strengths.

The frequencies of oscillation are given by $\Omega'_l=\Omega_l/\omega=|\textrm{Re}(i\lambda_l^{1\pm})|$ and $\Lambda_l'=\Lambda_l/\omega=|\textrm{Re}(i\lambda_l^{2\pm})|$, and read up to the first order of $\delta_l$ 
\begin{equation}
\Omega'_l=\Omega_0'+\frac{1}{\Omega_0'^3}
   \left(
     \frac{11\sigma_0'^2}{8}
     +\frac{7}{4\sigma_0'^2}
     +\frac{7}{8\sigma_0'^6}
   \right)\delta_l \hspace{1.5cm} \label{AppOsc''}
\end{equation}
and
\begin{eqnarray}
\hspace{-1.5cm} \Lambda'_l=\frac{1}{\Omega'_0}\sqrt{\frac{P}{2\sigma'_0}\left(
      5+\frac{3}{\sigma_0'^4}
    \right)\delta_l
  +\frac{1}{\Omega_0'^4}
  \left(
      -\frac{5\sigma_0'^4}{4}
      +\frac{45}{4}       
      +\frac{11}{4\sigma_0'^4}
      +\frac{7}{4\sigma_0'^8}
      +\frac{3}{2\sigma_0'^{12}}
   \right)\delta_l^2}. \label{AppOsc'}
\end{eqnarray}
These equations are the dimensionless counterpart of equations~(\ref{Osc''}) and (\ref{Osc'}) from the main text. The matrix $Q_l$ has $4$ different eigenvalues $\lambda_l^{1\pm}=\pm i\Omega'_l, \lambda_l^{2\pm}=\pm i\Lambda'_l$, where $\Omega'_l$ and $\Lambda'_l$ are positive real numbers if $P$ is positive or for small negative values of it. The associated eigenvectors read
\begin{equation}
X_{l}^{1\pm}=\left(\begin{array}{cccc}
\displaystyle -\frac{P}{2\sigma_0'^2\Omega_0'^2}\delta_l \\ \\
\displaystyle \mp\frac{i}{\Omega_0'}
  \left(
  2\sigma_0'+\frac{P}{2\sigma'^2_0}
  +\frac{P}{16\Omega_0'^4}
  \frac{[5\sigma'^8_0-82\sigma'^4_0-35]}{\sigma'^8_0}
\delta_l
\right)
 \\ \\
1 \\ \\
\displaystyle \mp \frac{i\Omega_0'}{\sigma_0'}
  \left(
    1-\frac{P}{\Omega'^4_0}\frac{[7\sigma'^4+5]}{\sigma'^6}\delta_l
  \right)
\end{array}\right) \label{B45}
\end{equation}
and
\begin{equation}
X_{l}^{2\pm}=\left(\begin{array}{cccc}
1 \\ \\
\displaystyle \pm 2i\frac{\Lambda'_l}{\delta_l}
  \left(
    1-\frac{P}{4\sigma'_0\Omega'^2_0}\delta_l
  \right)
\\ \\
\displaystyle \frac{P}{\sigma_0'^2\Omega_0'^2}\left\{1+
  \left[
    -\frac{1}{\sigma'^2_0\Omega_0'^2}
    +\frac{P}{2\sigma'_0\Omega_0'^4}
      \left(
        5+\frac{3}{\sigma'^4_0}
      \right)
  \right]
\delta_l\right\}
\\ \\
\displaystyle \mp i\frac{P\Lambda'_l}{\sigma_0'^3\Omega_0'^2}\left\{1+
  \left[
    -\frac{1}{\sigma'^2_0\Omega_0'^2}
    +\frac{P}{2\sigma'_0\Omega_0'^4}
      \left(
        5+\frac{3}{\sigma'^4_0}
      \right)
  \right]
\delta_l\right\}
\end{array}\right). \label{B46}
\end{equation}
The solution of the system~(\ref{LinearSystem}) is thus given by~(\ref{Xlmt}), so we have the following time dependences
\begin{eqnarray}
e^{Q_lt'}X_{l}^{1\pm}=e^{\lambda_l^{1\pm}t'}X_{l}^{1\pm}, \ \ \ \ 
e^{Q_lt'}X_{l}^{2\pm}=e^{\lambda_l^{2\pm}t'}X_{l}^{2\pm}, \label{solutionsystem}
\end{eqnarray}
Note that the eigenvectors~(\ref{B45}), (\ref{B46}) are not pure real vectors, thus in principle, these vectors and the solution~(\ref{solutionsystem}) have no physical meaning. But the first and third entries of both vectors~(\ref{B45}), (\ref{B46}) are real numbers, while the second and fourth entries are purely imaginary numbers. Then the vectors
\begin{equation}
\hspace{-0.7cm} \frac{X_{l}^{1+}+X_{l}^{1-}}{2}
, \ \ \ \ \frac{i(X_{l}^{1+}-X_{l}^{1-})}{2}
, \ \ \ \ \frac{X_{l}^{2+}+X_{l}^{2-}}{2}
\ \ \ \textrm{and} \ \ \ \ 
i\frac{(X_{l}^{2+}-X_{l}^{2-})}{2}
\end{equation}
are real vectors. We point out that the dynamics obtained with the combinations $i(X_{l}^{n+}-X_{l}^{n-})/2$, for $n=1,2$, turn out to be the same of that obtained with $(X_{l}^{n+}+X_{l}^{n-})/2$, for $n=1,2$, apart from a phase. Therefore, we consider from now only the latter ones.

With this we obtain
\begin{eqnarray}
\hspace{-1.7cm} X_{l}^1(t')=e^{Qt'}\frac{X_{l}^{1+}+X_{l}^{1-}}{2}&=\frac{e^{\lambda_l^1t'}X_{l}^{1+}+e^{-\lambda_l^1t'}X_{l}^{1-}}{2} 
\\
&=\cos(\Omega_l't')\frac{X_{l}^{1+}+X_{l}^{1-}}{2}+i\sin(\Omega_l't')\frac{(X_{l}^{1+}-X_{l}^{1-})}{2},
\nonumber
\end{eqnarray}
which reads explicitly
\begin{equation}
\hspace{-1cm} X_{l}^1(t')=\left(
\begin{array}{cccc}
\displaystyle -\frac{P}{2\sigma_0'^2\Omega_0'^2}\delta_l\cos(\Omega_l't') \\ \\
\displaystyle \frac{1}{\Omega_0'}
  \left(
  2\sigma_0'+\frac{P}{2\sigma'^2_0}
  +\frac{P}{16\Omega_0'^4}
  \frac{[5\sigma'^8_0-82\sigma'^4_0-35]}{\sigma'^8_0}
\delta_l
\right)\sin(\Omega_l't') \\ \\
\cos(\Omega_l't') \\ \\
\displaystyle \frac{\Omega_0'}{\sigma_0'}
  \left(
    1-\frac{P}{\Omega'^4_0}\frac{[7\sigma'^4+5]}{\sigma'^6}\delta_l
  \right)\sin(\Omega_l't')
\end{array}\right)
\end{equation}
with the initial condition
\begin{equation}
\hspace{-1cm}X_{l}^1(0)=\left(\begin{array}{cccc}
\displaystyle -\frac{P}{2\sigma_0'^2\Omega_0'^2}\delta_l \\ \\
0 \\ \\
1 \\ \\
0 \end{array}\right).
\end{equation}
Analogously, we also obtain
\begin{eqnarray}
\hspace{-1cm}X_{l}^2(t')=
\left(\begin{array}{cccc}
\cos(\Lambda_l't')\\ \\
\displaystyle -2\frac{\Lambda'_l}{\delta_l}
  \left(
    1-\frac{P}{4\sigma'_0\Omega'^2_0}\delta_l
  \right)\sin(\Lambda_l't') \\ \\
\displaystyle \frac{P}{\sigma_0'^2\Omega_0'^2}\left\{1+
  \left[
    -\frac{1}{\sigma'^2_0\Omega_0'^2}
    +\frac{P}{2\sigma'_0\Omega_0'^4}
      \left(
        5+\frac{3}{\sigma'^4_0}
      \right)
  \right]
\delta_l\right\}\cos(\Lambda_l't') \\ \\
\displaystyle \frac{P\Lambda'_l}{\sigma_0'^3\Omega_0'^2}\left\{1+
  \left[
    -\frac{1}{\sigma'^2_0\Omega_0'^2}
    +\frac{P}{2\sigma'_0\Omega_0'^4}
      \left(
        5+\frac{3}{\sigma'^4_0}
      \right)
  \right]
\delta_l\right\}\sin(\Lambda_l't')
\end{array}\right)
\end{eqnarray}
with the corresponding initial condition
\begin{eqnarray}
\hspace{-1cm}X_{l}^2(0)=\left(\begin{array}{cccc}
1 \\ \\
0 \\ \\
\displaystyle \frac{P}{\sigma_0'^2\Omega_0'^2}\left\{1+
  \left[
    -\frac{1}{\sigma'^2_0\Omega_0'^2}
    +\frac{P}{2\sigma'_0\Omega_0'^4}
      \left(
        5+\frac{3}{\sigma'^4_0}
      \right)
  \right]
\delta_l\right\} \\ \\
0 \end{array}\right).
\end{eqnarray}
Thus, performing at $t'=0$ a small perturbation given by
\begin{equation}
\delta\psi'_{lm}(0)=-C^1_{lm}\frac{P}{4\sigma_0'^2\Omega_0'^2}\delta_l\bar{Y}_{lm},
\ \ \ \ \ 
\delta\sigma'_{lm}(0)=C_{lm}^1\bar{Y}_{lm} ,
\label{AppPert''}
\end{equation}
where $C_{lm}^1$ is a proportionality constant, the evolution of this system is given by
\begin{eqnarray}
\hspace{-2.5cm}\delta\psi'_{lm}(t')=C_{lm}^1
  \Bigg\{
    -\frac{P}{4\sigma_0'^2\Omega_0'^2}\delta_l\cos(\Omega'_l t') \nonumber
   \\ +\frac{i}{\Omega_0'}
  \left(
  \sigma_0'+\frac{P}{4\sigma'^2_0}
  +\frac{P}{32\Omega_0'^4}
  \frac{[5\sigma'^8_0-82\sigma'^4_0-35]}{\sigma'^8_0}
\delta_l
\right)\sin(\Omega'_l t')
  \Bigg\}\bar{Y}_{lm}, \nonumber \\
\hspace{-2.5cm}\delta\sigma'_{lm}(t')=C_{lm}^1\cos(\Omega'_l t')\bar{Y}_{lm}, \ \ \ \ \ \label{AppDyn''} \\
\hspace{-2.5cm}\delta B'_{lm}(t')=C_{lm}^1\frac{\Omega_0'}{\sigma_0'}
  \left(
    1-\frac{P}{\Omega'^4_0}\frac{[7\sigma'^4+5]}{\sigma'^6}\delta_l
  \right)\sin(\Omega'_l t')\bar{Y}_{lm}. \nonumber
\end{eqnarray}
These equations are the dimensionless form of equations~(\ref{Dyn''}) in the main text.
But a small perturbation at $t'=0$ given by
\begin{eqnarray}
\hspace{-2.5cm}\delta\psi'_{lm}(0)=\frac{C_{lm}^2}{2}\bar{Y}_{lm}, \label{AppPert'} \\
\hspace{-2.5cm}\delta\sigma'_{lm}(0)=C_{lm}^2\frac{P}{\sigma_0'^2\Omega_0'^2}\left\{1+
  \left[
    -\frac{1}{\sigma'^2_0\Omega_0'^2}
    +\frac{P}{2\sigma'_0\Omega_0'^4}
      \left(
        5+\frac{3}{\sigma'^4_0}
      \right)
  \right]
\delta_l\right\}
\bar{Y}_{lm}, \nonumber
\end{eqnarray}
where $C_{lm}^2$ is also a proportionality constant, leads to the solution
\begin{eqnarray}
\hspace{-2.5cm}\delta\psi'_{lm}(t')=C_{lm}^2
  \left(
     \frac{1}{2}\cos(\Lambda'_l  t')
    -i\frac{\Lambda'_l}{\delta_l}
  \left(
    1-\frac{P}{4\sigma'_0\Omega'^2_0}\delta_l
  \right)\sin(\Lambda'_l  t')
  \right)\bar{Y}_{lm}, \nonumber \\
\hspace{-2.5cm}\delta\sigma'_{lm}(t')=C_{lm}^2\frac{P}{\sigma_0'^2\Omega_0'^2}\left\{1+
  \left[
    -\frac{1}{\sigma'^2_0\Omega_0'^2}
    +\frac{P}{2\sigma'_0\Omega_0'^4}
      \left(
        5+\frac{3}{\sigma'^4_0}
      \right)
  \right]
\delta_l\right\}\cos(\Lambda'_l t')
\bar{Y}_{lm},
 \ \ \ \ \  \label{AppDyn'}\\
\hspace{-2.5cm}\delta B'_{lm}(t')=C_{lm}^2\frac{P\Lambda'_l}{\sigma_0'^3\Omega_0'^2}\left\{1+
  \left[
    -\frac{1}{\sigma'^2_0\Omega_0'^2}
    +\frac{P}{2\sigma'_0\Omega_0'^4}
      \left(
        5+\frac{3}{\sigma'^4_0}
      \right)
  \right]
\delta_l\right\}\sin(\Lambda'_l  t')
\bar{Y}_{lm},  \nonumber
\end{eqnarray}
which are the dimensionless form of equations~(\ref{Dyn'}) in the main text.

A small perturbation of $\delta\sigma'_{lm}$ and of the real part of $\delta\psi'_{lm}$ which is not proportional to (\ref{AppPert''}) or to (\ref{AppPert'}) is a linear combination of them, thus the corresponding evolution of the state is given by a linear combination of~(\ref{AppDyn''}) and (\ref{AppDyn'}).

\end{document}